\documentclass[10pt,journal,compsoc,final]{IEEEtran}

\usepackage[frozencache=true]{minted}

\usepackage[font=scriptsize]{caption}
\usepackage[font=scriptsize]{subcaption}

\usepackage{moresize}
\usepackage{comment}
\usepackage{ifdraft}

\usepackage{listings}

\DeclareCaptionSubType{listing} 

\usepackage[hidelinks]{hyperref} 

\fvset{linenos, fontsize=\scriptsize, frame=lines, numbersep=1pt, xleftmargin=.75em}
\usepackage{mathtools} 

\DeclarePairedDelimiter{\floor}{\lfloor}{\rfloor}
\usepackage{multirow} 
\usepackage{enumitem} 
\usepackage{color}
\usepackage{hhline} 
\usepackage{array}
\usepackage{makecell,cellspace}
\usepackage{fontawesome} 
\usepackage{amssymb} 
\usepackage{pifont} 
\usepackage{moresize}
\newcommand{\xmark}{\ding{55}}%
\usepackage[table]{xcolor} 
\usepackage[per-mode=symbol]{siunitx}
\DeclareSIUnit \bit {bit}
\DeclareSIUnit \byte {Byte}
\DeclareSIUnit \cycle {cycle}
\DeclareSIUnit \cycles {cycles}
\DeclareSIUnit \hz {Hz}
\DeclareSIUnit \op {Op}
\DeclareSIUnit \operand {operand}
\DeclareSIUnit \operands {operands}
\DeclareSIUnit \transfer {T}
\DeclareSIUnit \cell {cell}

\usepackage{booktabs} 

\newcommand{\figureref}[1]{\hyperref[fig:#1]{Fig.~\ref{fig:#1}}}
\newcommand{\tableref}[1]{\hyperref[tab:#1]{Tab.~\ref{tab:#1}}}
\newcommand{\listingref}[1]{\hyperref[lst:#1]{Lst.~\ref{lst:#1}}}
\newcommand{\equref}[1]{\hyperref[eq:#1]{Eq.~\ref{eq:#1}}}
\newcommand{\secref}[1]{\hyperref[sec:#1]{Sec.~\ref{sec:#1}}}
\newcommand{\coderef}[1]{line~\ref{code:#1}}

\usepackage{bm}
\newcommand{\matr}[1]{\bm{#1}}

\newcommand{\secsymbol}{{\S}}
\newcommand{\seclink}[1]{\hyperref[sec:#1]{\secsymbol\ref{sec:#1}}}

\usepackage{varwidth}
\makeatletter
\newcommand{\breakcaption}{\@dblarg\emit@breakcaption}
\long\def\emit@breakcaption[#1]#2{%
  \expandafter\caption\expandafter[\expandafter\emit@removeafter#1\\\@nil]{%
    \begin{varwidth}[t]{\columnwidth-\widthof{\figurename\space\thefigure:\space}}
    #2
    \end{varwidth}%
  }%
}
\def\emit@removeafter#1\\#2\@nil{#1}
\makeatother

\usepackage[colorinlistoftodos,prependcaption,textsize=tiny,textwidth=0.7cm]{todonotes}
\usepackage{xargs}
\newcommandx{\unsure}[2][1=]{\todo[linecolor=red,backgroundcolor=red!25,bordercolor=red,#1]{#2}}
\newcommandx{\change}[2][1=]{\todo[linecolor=blue,backgroundcolor=blue!25,bordercolor=blue,#1]{#2}}
\newcommandx{\info}[2][1=]{\todo[linecolor=OliveGreen,backgroundcolor=OliveGreen!25,bordercolor=OliveGreen,#1]{#2}}
\newcommandx{\improvement}[2][1=]{\todo[linecolor=Plum,backgroundcolor=Plum!25,bordercolor=Plum,#1]{#2}}

\usepackage{skull}
\usepackage{tikzsymbols}

\usepackage{inconsolata}

\newminted{c}{
    linenos,
    escapeinside=||,
    fontsize=\tt\scriptsize,
    numbersep=3pt,
    xleftmargin=0.75em,
    frame=lines,
    tabsize=2,
    breaklines=true
}
\newif\ifFirstMintedPart
\makeatletter
\newenvironment{mintedBlock}{%
    \par
    \medskip
    \begingroup
        \setlength{\parskip}{0pt}%
        \setlength{\baselineskip}{0pt}%
        \setlength{\lineskip}{0pt}%
        \let\originalVspace=\vspace
        \renewcommand{\vspace}{\@ifnextchar*\@gobbletwo\@gobble}%
        \setlength{\fboxsep}{0pt}%
        \FirstMintedParttrue
        \noindent\FancyVerbRuleColor{\vrule \@width\linewidth \@height\FV@FrameRule}%
        \originalVspace{2pt}
        \par
}{%
        \par
        \originalVspace{2pt}
        \noindent\FancyVerbRuleColor{\vrule \@width\linewidth \@height\FV@FrameRule}%
        \par
    \endgroup
    \medskip
}
\makeatother
\newenvironment{cpart}[1]{%
    \VerbatimEnvironment
    \ifFirstMintedPart
        \newcommand\currentLineNumber{firstnumber=1}%
    \else
        \newcommand\currentLineNumber{firstnumber=last}%
    \fi
    \newcommand{\beginCCode}{\begin{ccode*}}%
    \expandafter\beginCCode\expandafter{%
        \currentLineNumber,
        frame=none,
        bgcolor=white,
        breaklines, 
        #1
    }%
}{%
    \end{ccode*}%
    \global\FirstMintedPartfalse
}
\newcommand{\mintedalt}{black!10}%

\newcommand{\one}{\ding{182}}
\newcommand{\two}{\ding{183}}
\newcommand{\three}{\ding{184}}
\newcommand{\four}{\ding{185}}

\begin{document}

\title{\vspace{-0.5em}Transformations of High-Level Synthesis Codes\\for High-Performance Computing\vspace{-0.5em}}

\author{%
\hspace{.02\textwidth}%
\begin{minipage}{.25\textwidth}
  \centering
  Johannes~de~Fine~Licht\\
  definelicht@inf.ethz.ch
\end{minipage}%
\begin{minipage}{.29\textwidth}
  \centering
  Maciej~Besta\\
  maciej.besta@inf.ethz.ch
\end{minipage}%
\begin{minipage}{.19\textwidth}
  \centering
  Simon~Meierhans\\
  mesimon@ethz.ch
\end{minipage}%
\begin{minipage}{.23\textwidth}
  \centering
  Torsten~Hoefler\\
  htor@inf.ethz.ch
\end{minipage}%
\hspace{.02\textwidth}\\[2mm]
Department of Computer Science, ETH Zurich\vspace{-12pt}}


\IEEEtitleabstractindextext{%
\begin{abstract}
%
%
Spatial computing architectures promise a major stride in performance and energy
efficiency over the traditional load/store devices currently employed in large
scale computing systems. 
The adoption of high-level synthesis (HLS) from languages such as C++ and OpenCL
has greatly increased programmer productivity when designing for such platforms.
While this has enabled a wider audience to target spatial computing
architectures, the optimization principles known from traditional software
design are no longer sufficient to implement high-performance codes, due to
fundamentally distinct aspects of hardware design, such as programming for deep
pipelines, distributed memory resources, and scalable routing.
%
%
To alleviate this, we present a collection of optimizing transformations for
HLS, targeting scalable and efficient architectures for high-performance
computing (HPC) applications.
We systematically identify \emph{classes} of transformations (pipelining,
scalability, and memory), the \emph{characteristics} of their effect on the HLS
code and the resulting hardware (e.g., increasing data reuse or resource
consumption), and the \emph{objectives} that each transformation can target
(e.g., resolve interface contention, or increase parallelism).
We show how these can be used to efficiently exploit pipelining, on-chip
distributed fast memory, and on-chip dataflow, allowing for massively parallel
architectures. 
To quantify the effect of various transformations, we cover the optimization
process of a sample set of HPC kernels, provided as open source reference codes. 
%
%
We aim to establish a common toolbox to guide both performance engineers and
compiler engineers in tapping into the performance potential offered by spatial
computing architectures using HLS.
\end{abstract}
}

\maketitle
\vspace{-2em} 

\newcommand{\thumbsup}{\faThumbsOUp}
\newcommand{\thumbsdown}{\faThumbsDown}
\newcommand{\moregood}{{\thumbsup}}
\newcommand{\mmoregood}{{\thumbsup\textbf{!}}}
\newcommand{\morebad}{{\thumbsdown}}
\newcommand{\mmorebad}{{\thumbsdown\textbf{!}}}
\newcommand{\lessgood}{\thumbsup}
\newcommand{\lessbad}{\thumbsdown}
\newcommand{\llessbad}{{\thumbsdown\textbf{!}}}
\newcommand{\same}{--}
\newcommand{\samegood}{\thumbsup}
\newcommand{\sometimes}{$\sim$}
\newcommand{\yesgood}{\thumbsup}
\newcommand{\yesbad}{\thumbsdown}
\newcommand{\yes}{\xmark}
\newcommand{\objective}{\faHandOLeft}
\newcolumntype{x}{>{\centering\let\newline\\\arraybackslash\hspace{0pt}}p{.047\columnwidth}}

\section{Introduction}
\vspace{-0.25em} 
\label{sec:introduction}
\label{sec:motivation}

Since the end of Dennard scaling, when the power consumption of digital circuits
stopped scaling with their size, compute devices have become increasingly
limited by their power consumption~\cite{hitting_the_power_wall}. 
In fact, shrinking the feature size even \emph{increases} the loss in the metal
layers of modern microchips. 
Today's load/store architectures suffer mostly from the cost of data movement
and addressing general purpose registers and
cache~\cite{computings_energy_problem}.
Other approaches, such as dataflow architectures, have not been widely
successful, due to the varying granularity of
applications~\cite{second_opinion_on_dataflow}.
However, \emph{application-specific} dataflow can be used to lay out as
registers and on-chip memory to fit the \emph{specific structure}
of the computation, and thereby minimize data movement. 

Reconfigurable architectures, such as FPGAs, can be used to implement
application-specific dataflow~\cite{wheres_the_beef, platform_comparison,
rng_comparison}, but are hard to program~\cite{fpga_for_the_masses}, as
traditional hardware design languages, such as VHDL or Verilog, do not benefit
from the rich set of software engineering techniques that improve programmer
productivity and code reliability. 
For these reasons, both hardware and software communities are embracing
high-level synthesis (HLS) tools~\cite{hls_past_present_future,hls_for_fpgas},
enabling hardware development using procedural languages.

HLS bridges the gap between hardware and software development, and enables basic
performance portability implemented in the compilation system. 
For example, HLS programmers do not have to worry about how exactly a floating
point operation, a bus protocol, or a DRAM controller is implemented on the
target hardware. 
%
%
Numerous HLS systems~\cite{hls_survey_nane, hls_survey_meeus} synthesize
hardware designs from C/C++ \cite{autopilot, intel_hls, legup, catapult_c,
bambu, dwarv}, OpenCL \cite{opencl_owaida, altera_opencl} and other high-level
languages \cite{bluespec, lime, liquid_metal, esterel, streamsc},
providing a viable path for software and hardware communities to meet and
address each other's concerns.


For many applications, computational performance is a primary goal, which is
achieved through careful tuning by specialized performance engineers using
well-understood optimizing transformations when targeting
CPU~\cite{compiler_transformations} and GPU~\cite{gpu_optimization}
architectures.
\textbf{For HLS, a comparable collection of guidelines and principles for code
optimization is yet to be established.}
Optimizing codes for hardware is drastically different from optimizing codes for
software. 
In fact, the optimization space is \emph{larger}, as it contains most known
software optimizations, in addition to HLS-specific transformations that let
programmers manipulate the underlying hardware architecture. 
To make matters worse, the low clock frequency, lack of cache, and fine-grained
configurability, means that \emph{naive} HLS codes typically perform poorly
compared to naive software codes, and must be transformed considerably before
the advantages of specialized hardware can be exploited. 
\textbf{Thus, the established set of traditional transformations is
insufficient, as it does not consider aspects of optimized hardware design,
such as pipelining and decentralized fast memory.}


In this work, we survey and define a set of key transformations that optimizing
compilers or performance engineers can apply to improve the performance of
hardware layouts generated from HLS codes. 
This set unions transformations extracted from previous work, where they were
applied either explicitly or implicitly, with additional techniques that fill in
gaps to maximize completeness.
%
%
We characterize and categorize transformations, allowing performance engineers
to easily look up those relevant to improving their HLS code, based on the
problems and bottlenecks currently present.
%
%
The transformations have been verified to apply to both the Intel OpenCL and
Xilinx Vivado HLS toolflows, but are expected to translate to any pragma-based
imperative HLS tool.

\noindent In addition to identifying previous work that apply one or more of the
transformations defined here, we describe and publish a set of end-to-end
``hands-on'' examples, optimized from naive HLS codes into high performance
implementations. This includes a stencil code, matrix multiplication, and the
N-body problem, all available on github.
The optimized codes exhibit dramatic cumulative speedups of up to
$\SI{29950}{}{\times}$ relative to their respective naive starting points,
showing the crucial necessity of hardware-aware transformations, which are not
performed automatically by today's HLS compilers.
As FPGAs are currently the only platforms commonly targeted by HLS tools in the
HPC domain, transformations are discussed and evaluated in this context.
Evaluating FPGA performance in comparison to other platforms is out of scope of
this work.
\textbf{Our work provides a set of guidelines and a cheat~sheet for optimizing
high-performance codes for reconfigurable architectures, guiding both
performance engineers and compiler developers to efficiently exploit these
devices.}

\begin{table}
\centering
\newcommand{\transline}{\hhline{~----------}}
\newcommand{\transsepline}{\hhline{~==========}}
\setlength\tabcolsep{0.75pt}
\ssmall
\sf
\newcolumntype{P}[1]{>{\centering\arraybackslash}p{#1}}
\begin{tabular}{r l l | c c c c c c c c | c c c c c c c}
\toprule

\multicolumn{3}{c}{\textbf{Transformations}} &
\multicolumn{8}{c}{\textbf{Characteristics}} &
\multicolumn{7}{c}{\textbf{Objectives}} \\
\multicolumn{3}{l}{} & \textbf{PL} & \textbf{RE} & \textbf{PA} & \textbf{ME} & \textbf{RS} & \textbf{RT} & \textbf{SC} & \textbf{CC} &
                       \textbf{LD} & \textbf{RE} & \textbf{CU} & \textbf{BW} & \textbf{PL} & \textbf{RT} & \textbf{RS} \\
\midrule
%
\parbox[t]{2mm}{\multirow{7}{*}{\rotatebox[origin=c]{90}{\ssmall\textbf{Pipelining}}}}
& Accumulation interleaving & \seclink{interleaving}    & \moregood   & \same       & \same       & \sometimes & \morebad    & \same      & \yesbad    & \sometimes 
                            & \objective                & \same       & \same       & \same      & \same       & \same      & \same \\
& Delay buffering           & \seclink{fifo_buffering}  & \moregood   & \moregood   & (\moregood) & \moregood  & \morebad    & (\lessbad) & \lessbad   & \lessbad
                            & \objective                & \objective  & \same       & \same      & \same       & \same      & \same \\ 
& Random access buffering   & \seclink{ram_buffering}   & \moregood   & \moregood   & (\moregood) & \moregood  & \morebad    & \lessbad   & \lessbad   & \lessbad
                            & \objective                & \objective  & \same       & \objective & \same       & \same      & \same \\
& Pipelined loop fusion     & \seclink{fusion}          & \moregood   & (\moregood) & \same       & \sometimes & \sometimes  & (\lessbad) & \same      & \lessbad 
                            & \same                     & \same       & \same       & \same      & \objective  & \same      & \same \\
& Pipelined loop switching  & \seclink{loop_switching}  & \moregood   & (\moregood) & \same       & \sometimes & \sometimes  & (\lessbad) & \same      & \sometimes 
                            & \same                     & \same       & \same       & \same      & \objective  & \same      & \objective \\
& Pipelined loop flattening & \seclink{flattening}      & \moregood   & \same       & \same       & \moregood  & \sometimes  & \sometimes & \same      & \lessbad
                            & \same                     & \same       & \same       & \same      & \objective  & \same      & \same \\
& Inlining                  & \seclink{inlining}        & \moregood   & \same       & (\moregood) & \same      & (\morebad)  & \same      & \same      & \samegood
                            & \objective                & \same       & \same       & \same      & \same       & \same      & \same \\
%
\midrule
\parbox[t]{2mm}{\multirow{4}{*}{\rotatebox[origin=c]{90}{\ssmall\textbf{Scaling}}}}
& Horizontal unrolling    & \seclink{vectorization}   & \same       & (\moregood) & \moregood   & \moregood  & \morebad   & \lessbad   & \yesbad    & (\lessbad)
                          & \same                     & \same       & \objective  & \same      & \same      & \same      & \same \\
& Vertical unrolling      & \seclink{replication}     & \same       & \mmoregood  & \mmoregood  & \same      & \mmorebad  & \llessbad  & \yesbad    & \lessbad   
                          & \same                     & \objective  & \objective  & \same      & \same      & \same      & \same \\
& Dataflow                & \seclink{dataflow}        & \same       & \same       & (\moregood) & \same      & (\morebad) & \mmoregood & \same      & \moregood 
                          & \same                     & \objective  & \same       & \same      & \objective & \objective & \same \\
& Tiling                  & \seclink{tiling}          & \same       & \moregood   & \same       & \sometimes & \morebad   & \sometimes & \yesbad    & \lessbad  
                          & \objective                & \objective  & \same       & \same      & \same      & \objective & \objective \\
\midrule
\parbox[t]{2mm}{\multirow{4}{*}{\rotatebox[origin=c]{90}{\ssmall\textbf{Memory}}}}
& Mem. access extraction & \seclink{extraction}       & (\moregood) & \same       & \same       & \moregood  & \morebad   & \moregood  & \same      & \lessbad
                         & \objective                 & \same       & \same       & \objective & \same      & \same      & \same \\
& Mem. buffering         & \seclink{memory_buffering} & \same       & \same       & \same       & \moregood  & \morebad   & \same      & \same      & \lessbad
                         & \same                      & \same       & \same       & \objective & \same      & \same      & \same \\
& Mem. striping          & \seclink{striping}         & \same       & \same       & \same       & \moregood  & \morebad   & \lessbad   & \same      & \lessbad
                         & \same                      & \same       & \same       & \objective & \same      & \same      & \same \\
& Type demotion          & \seclink{type_demotion}    & \same       & \same       & \same       & \moregood  & \lessgood  & \moregood  & \same      & \samegood
                         & \same                      & \same       & \same       & \objective & \same      & \same      & \objective \\
\bottomrule
\end{tabular}
\vspace{-1em}
\caption{%
Overview of \textbf{transformations}, the \textbf{characteristics} of their
effect on the HLS code and the resulting hardware, and the \textbf{objectives}
that they can target.
The center group of column marks the following transformation characteristics:
(\textbf{PL}) \emph{enables pipelining};
(\textbf{RE}) \emph{increases data reuse}, i.e., increases the arithmetic
intensity of the code;
(\textbf{PA}) \emph{increases or exposes more parallelism};
(\textbf{ME}) \emph{optimizes memory accesses};
(\textbf{RS}) \emph{does \textbf{not} significantly increase resource
consumption};
(\textbf{RT}) \emph{does \textbf{not} significantly impair routing}, i.e.,
does \textbf{not} potentially reduce maximum frequency or prevent the design
from being routed altogether;
(\textbf{SC}) \emph{does \textbf{not} change the schedule of loop nests},
e.g., by introducing more loops; and
(\textbf{CC}) \emph{does \textbf{not} significantly increase code complexity}.
The symbols have the following meaning:
``\same'': no effect,
\mbox{``\moregood''}: positive effect,
``\mmoregood'': very positive effect,
``(\moregood)'': small or situational positive effect,
``\morebad'': negative effect,
``\mmorebad'': very negative effect,
``(\morebad)'': small or situational negative effect,
``$\sim$'': positive or negative effect can occur, depending on the context.
\quad The right group of columns marks the following \textbf{objectives} that
can be targeted by transformations:
(\textbf{LD}) \emph{resolve loop-carried dependencies}, due to inter-iteration
dependencies or resource contention; 
(\textbf{RE}) \emph{increase data reuse}; 
(\textbf{CU}) \emph{increase parallelism}; 
(\textbf{BW}) \emph{increase memory bandwidth utilization};
(\textbf{PL}) \emph{reduce pipelining overhead};
(\textbf{RT}) \emph{improve routing results};
(\textbf{RS}) \emph{reduce resource utilization}. 
}
\vspace{-1em}
\label{tab:hls_transformations}
\end{table}

\vspace{-0.75em} 
\subsection{From Imperative Code to Hardware}
\vspace{-0.25em}
\label{sec:hls_procedure}
\label{sec:imperative_code_to_hardware}

Before diving into transformations, it is useful to form an intuition of the
major stages of the source-to-hardware stack, to understand how they are
influenced by the HLS code: 
\\[1mm]
\one{} \textbf{High-level synthesis} converts a pragma-assisted procedural
description (C++, OpenCL) to a functionally equivalent behavioral description
(Verilog, VHDL). This requires mapping variables and operations to corresponding
constructs, then scheduling operations according to their inter-dependencies.
The dependency analysis is concerned with creating a hardware mapping such that
the throughput requirements are satisfied, which for pipelined sections require
the circuit to accept a new input every cycle. Coarse-grained control flow is
implemented with state machines, while computations and fine-grained control
flow are organized in (predicated) pipelines. 
\two{} \textbf{Hardware synthesis} maps the register-level circuit description to
components and wires present on the specific target \emph{architecture}. At this
stage and onwards, the procedure is both vendor and architecture specific. 
\three{} \textbf{Place and route} maps the logical circuit description to
concrete locations on the target \emph{device}, by performing a lengthy
heuristic-based optimization that attempts to minimize the length of the
longest wire and the total wire length. The longest propagation time between
two registers including the logic between them (i.e., the critical path of the
circuit), will determine the maximum obtainable frequency.
\four{} \textbf{Bitstream generation} translates the final circuit description
to a binary format used to configure the device. 
\\[1mm]
Most effort invested by an HLS programmer lies in guiding the scheduling process
in \one{} to implement deep, efficient pipelines, but \two{} is considered when
choosing data types and buffer sizes, and \three{} can ultimately bottleneck
applications once the desired parallelism has been achieved, requiring the
developer to adapt their code to aid this process.

\vspace{-0.75em} 
\subsection{Key Transformations for High-Level Synthesis}
\vspace{-0.25em}
\label{sec:contributions}
\label{sec:transformation_list}

This work identifies a set of optimizing transformations that are essential to
designing scalable and efficient hardware kernels in HLS\@.
%
%
An overview given in \tableref{hls_transformations}.
We divide the transformations into three major classes: \textbf{pipelining}
transformations, that enable or improve the potential for pipelining
computations; \textbf{scaling} transformations that increase or expose
additional parallelism; and \textbf{memory} enhancing transformations, which
increase memory utilization and efficiency.
Each transformation is further classified according to a number of
characteristic effects on the HLS source code, and on the resulting hardware
architecture (central columns).
To serve as a cheat sheet, the table furthermore lists common \emph{objectives}
targeted by HLS programmers, and maps them to relevant HLS transformations
(rightmost columns).
Characteristics and objectives are discussed in detail in relevant
transformation sections.

Throughout this work, we will show how each transformation is applied manually
by a performance engineer by directly modifying the source code, giving examples
before and after it is applied.  However, many transformations are also amenable
to automation in an optimizing compiler.  

\vspace{-0.75em} 
\subsection{The Importance of Pipelining}
\vspace{-0.25em} 
\label{sec:pipelining}

\begin{figure}[h]
  \centering
  \includegraphics[height=5em]{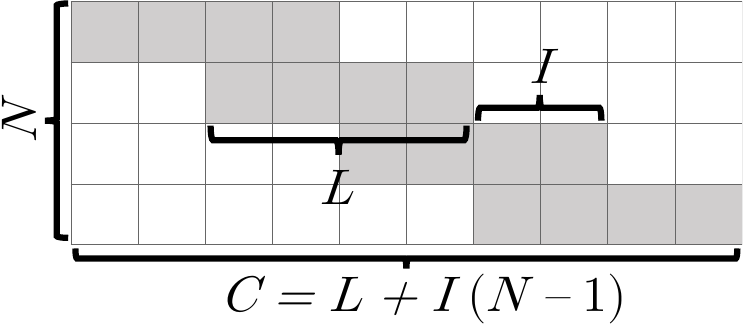}
  \vspace{-0.5em}
  \caption{Pipeline characteristics.}
  \label{fig:pipeline_characteristics}
\end{figure}

Pipelining is essential to efficient hardware architectures, as expensive
instruction decoding and data movement between memory, caches and registers can
be avoided, by sending data directly from one computational unit to the next. We
attribute two primary characteristics to pipelines:
\begin{itemize}[leftmargin=*]
  \item \textbf{Latency} ($L$): the number of cycles it takes for an input to
  propagate through the pipeline and arrive at the exit, i.e., the number of
  \textbf{pipeline stages}.
  \item \textbf{Initiation interval} or \textbf{gap} ($I$): the number of cycles
  that must pass before a new input can be accepted to the pipeline. A perfect
  pipeline has $I{=}\SI{1}{\cycle}$, as this is required to keep all pipeline
  stages busy. Consequently, the initiation interval can often be considered the
  \emph{inverse throughput} of the pipeline; e.g., $I{=}\SI{2}{\cycles}$ implies
  that the pipeline stalls every second cycle, reducing the throughput of
  \emph{all} pipelines stages by a factor of $\frac{1}{2}$.
\end{itemize}
To quantify the importance of pipelining in HLS, we consider the number of
cycles $C$ it takes to execute a pipeline with latency $L$ (both in
$[\si{\cycles}]$), taking $N$ inputs, with an initiation interval of
$I\;[\si{cycles}]$. Assuming a reliable producer and consumer at either end, we
have:
\begin{align}
  C = L + I \cdot (N - 1)\;[\si{\cycles}]\text{.} 
  \label{eq:pipeline}
\end{align}
This is shown in \figureref{pipeline_characteristics}. The time to execute all
$N$ iterations with clock rate $f\;[\si{\cycles\per\second}]$ of this pipeline
is thus $C/f$.

\noindent For two pipelines in sequence that both consume and produce $N$
elements, the latency is additive, while the initiation interval is decided by
the ``slowest'' actor:
\begin{align*}
  C_0 + C_1 = (L_0 + L_1) + \text{max}(I_0, I_1) \cdot (N - 1) 
\end{align*}
When $I_0{=}I_1$ this corresponds to a single, deeper pipeline. For large $N$,
the latencies are negligible, so this deeper pipeline increases pipeline
parallelism by adding more computations \emph{without increasing the runtime};
and without introducing additional off-chip memory traffic.
\textbf{We are thus interested in building deep, perfect pipelines to maximize
performance and minimize off-chip data movement.}

\vspace{-1em} 
\subsection{Optimization Goals}
\vspace{-0.25em} 
\label{sec:optimization_goal}

We organize the remainder of this work according to three overarching
optimization goals, corresponding to the three categories marked in
\tableref{hls_transformations}:
\begin{itemize}[leftmargin=*]
  \item \textbf{Enable pipelining} (\secref{pipeline_enabling_transformations}):
  For compute bound codes, achieve $I{=}\SI{1}{\cycle}$ for all essential
  compute components, to ensure that all pipelines run at maximum throughput.
  For memory bound codes, guarantee that memory is always consumed at line rate.
  \label{item:perfect_pipelining}
  \item \textbf{Scaling/folding} (\secref{scalability}): Fold the total number
  of iterations $N$ by scaling up the parallelism of the design to consume more
  elements per cycle, thus cutting the total number of cycles required to
  execute the program.  \label{item:scaling_folding}
  \item \textbf{Memory efficiency} (\secref{memory_transformations}): Saturate
  pipelines with data from memory to avoid stalls in compute logic. For memory
  bound codes, maximize bandwidth utilization.  \label{item:saturation}
\end{itemize}
\secref{compiler_transformations} covers the relationship between well-known
software optimizations and HLS, and accounts for which of these apply directly
to HLS code. \secref{evaluation} shows the effect of transformations on a
selection of kernels, \secref{related_work} presents related work, and we
conclude in \secref{conclusion}.


\vspace{-1.25em} 
\section{Pipeline-Enabling Transformations}
\vspace{-0.25em} 
\label{sec:pipeline_enabling_transformations}

\begin{figure}[h]
  \vspace{-1em} 
  \begin{minipage}{.5\columnwidth}
    \centering
    \includegraphics[height=8em]{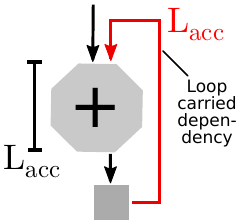}
    \vspace{-0.5em} 
    \caption{Loop-carried dependency.}
    \label{fig:loop_carried_dependency}
  \end{minipage}\hfill%
  \begin{minipage}{.5\columnwidth}
    \centering
    \includegraphics[height=8em]{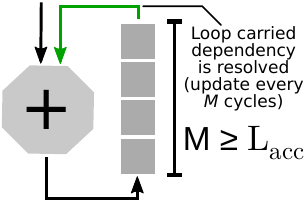}
    \vspace{-0.5em} 
    \caption{Buffered accumulation.}
    \label{fig:buffered_accumulation}
  \end{minipage}
  \vspace{-1em}
\end{figure}

As a crucial first step for any HLS code, we cover detecting and resolving
issues that prevent pipelining of computations.  When analyzing a basic block of
a program, the HLS tool determines the dependencies between computations, and
pipelines operations accordingly to achieve the target initiation interval.
There are two classes of problems that hinder pipelining of a given loop: 
\begin{enumerate}[leftmargin=*]
  \item \textbf{Loop-carried dependencies} (inter-iteration): an iteration of a
  pipelined loop depends on a result produced by a previous iteration, which
  takes multiple cycles to complete (i.e., has multiple internal pipeline
  stages). If the latency of the operations producing this result is $L$, the
  minimum initiation interval of the pipeline will be $L$. This is a common
  scenario when accumulating into a single register (see
  \figureref{loop_carried_dependency}), in cases where the accumulation
  operation takes $L_\text{acc}{>}1$ cycles.
  \label{item:loop_carried_dependency} 
  \item \textbf{Interface contention} (intra-iteration): a hardware resource
  with limited ports is accessed multiple times in the same iteration of the
  loop.  This could be a FIFO queue or RAM that only allows a single read and
  write per cycle, or an interface to external memory, which only supports
  sending/serving one request per cycle. 
\end{enumerate}
For each of the following transformations, we will give examples of programs
exhibiting properties that prevent them from being pipelined, and how the
transformation can resolve this.
All examples use C++ syntax, which allows classes (e.g., ``\texttt{FIFO}''
buffer objects) and templating. We perform pipelining and unrolling using pragma
directives, where loop-oriented pragmas always refer to the \emph{following
loop/scope}, which is the convention used by Intel/Altera HLS tools (as opposed
to applying to \emph{current} scope, which is the convention for Xilinx HLS
tools).

\vspace{-1em} 
\subsection{Accumulation Interleaving}
\vspace{-0.25em}
\label{sec:interleaving}

For multi-dimensional iteration spaces, \emph{loop-carried dependencies} can 
often be resolved by reordering and/or interleaving nested loops, keeping state
for multiple concurrent accumulations.
We distinguish between four approaches to interleaving accumulation, covered
below.

\begin{listing}
  \begin{subfigure}{\columnwidth}
    \begin{minipage}{.58\columnwidth}
      \begin{mintedBlock}
        \begin{cpart}{}
for (int n = 0; n < N; ++n)
  for (int m = 0; m < M; ++m) {
    double acc = C[n][m];
    #pragma PIPELINE
    for (int k = 0; k < K; ++k)
        \end{cpart}
        \begin{cpart}{bgcolor=\mintedalt}
      acc += A[n][k] * B[k][m]; |$\label{code:acc}$|
        \end{cpart}
        \begin{cpart}{}
    C[n][m] = acc; }
        \end{cpart}
      \end{mintedBlock}
    \end{minipage}\hfill%
    \begin{minipage}{.4\columnwidth}
      \flushright
      \includegraphics[height=7em]{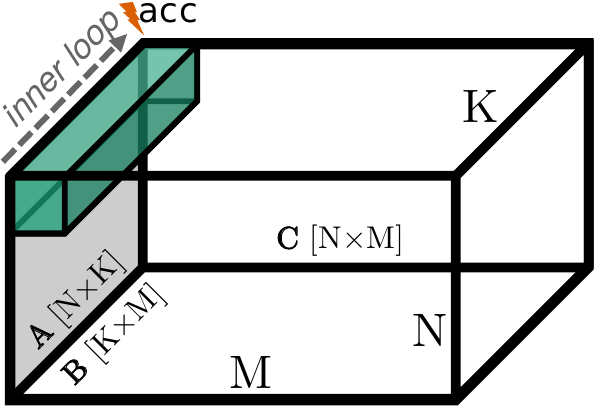}
    \end{minipage}
    \vspace{-1em}
    \caption{Naive implementation of general matrix multiplication $C {=} A B
    {+} C$.}
    \vspace{-0.25em}
    \label{lst:gemm_naive}
  \end{subfigure}
  \vspace{0.75em}
  \begin{subfigure}{\columnwidth}
    \begin{minipage}{.6\textwidth}
      \begin{mintedBlock}
        \begin{cpart}{}
for (int n = 0; n < N; ++n) {
  double acc[M]; // Uninitialized |$\label{code:acc_buffer}$|
  for (int k = 0; k < K; ++k)
    double a = A[n][k]; // Only read once
    #pragma PIPELINE
    for (int m = 0; m < M; ++m) { |$\label{code:gemm_p_loop}$|
      double prev = (k == 0) ? C[n][m]
                             : acc[m]; |$\label{code:acc_reset}$|
        \end{cpart}
        \begin{cpart}{bgcolor=\mintedalt}
      acc[m] = prev + a * B[k][m]; } |$\label{code:acc_write}$|
        \end{cpart}
        \begin{cpart}{}
  for (int m = 0; m < M; ++m) // Write 
    C[n][m] = acc[m]; }       // out 
        \end{cpart}
      \end{mintedBlock}
    \end{minipage}\hfill%
    \begin{minipage}{.4\textwidth}
      \flushright
      \includegraphics[height=7em]{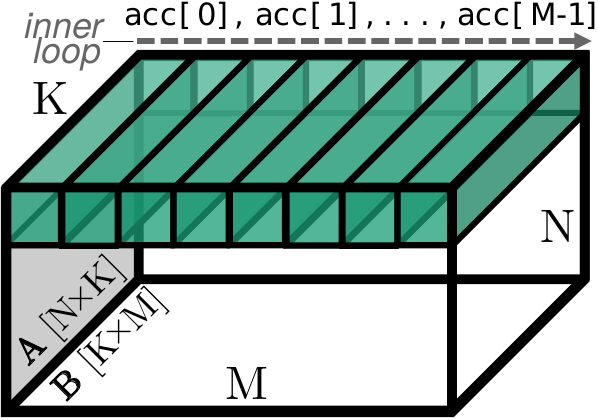}
    \end{minipage}
    \vspace{-1em}
    \caption{Transposed iteration space, same location written every $M$ cycles.}
    \vspace{-0.75em}
    \label{lst:gemm_transposed}
  \end{subfigure}
  \begin{subfigure}{\columnwidth}
    \begin{minipage}{.6\textwidth}
      \begin{mintedBlock}
        \begin{cpart}{}
for (int n = 0; n < N; ++n)
  for (int m = 0; m < M/T; ++m) {
    double acc[T]; // Tiles of size T 
    for (int k = 0; k < K; ++k)
      double a = A[n][k]; // M/T reads 
      #pragma PIPELINE
      for (int t = 0; t < T; ++t) {
        double prev = (k == 0) ?
            C[n][m*T+t] : acc[t];
        \end{cpart}
        \begin{cpart}{bgcolor=\mintedalt}
         acc[t] = prev + a * B[k][m*T+t]; }
        \end{cpart}
        \begin{cpart}{}
    for (int t = 0; t < T; ++t) // Write 
      C[n][m*T+t] = acc[t]; }   // out
        \end{cpart}
      \end{mintedBlock}
    \end{minipage}\hfill%
    \begin{minipage}{.4\textwidth}
      \flushright
      \includegraphics[height=7em]{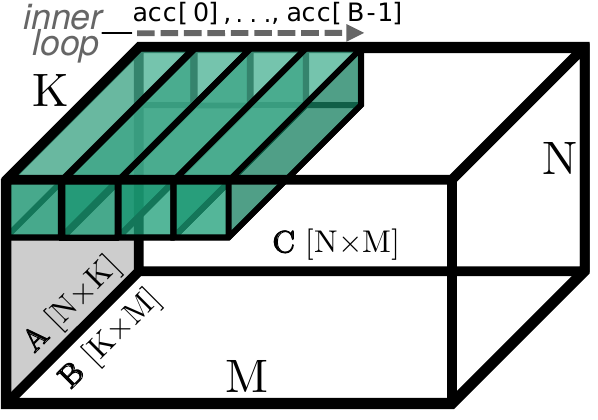}
    \end{minipage}
    \vspace{-1em}
    \caption{Tiled iteration space, same location written every $T$ cycles.}
    \label{lst:gemm_tiled}
  \end{subfigure}
  \vspace{-0.5em}
  \caption{Interleave accumulations to remove loop-carried dependency.}
  \label{lst:gemm}
\end{listing}

\vspace{-0.5em} 
\subsubsection{Full Transposition}
\vspace{-0.25em} 
\label{sec:transposition}
\label{sec:loop_interchange}

When a loop-carried dependency is encountered in a loop nest, it can be
beneficial to reorder the loops, thereby fully transposing the iteration space.
This typically also has a significant impact on the program's memory access
pattern, which can benefit/impair the program beyond resolving a loop-carried
dependency. 

Consider the matrix multiplication code in \listingref{gemm_naive}, computing
$\matr{C} = \matr{A} \cdot \matr{B} + \matr{C}$, with matrix dimensions $N$,
$K$, and $M$. The inner loop $k\in K$ accumulates into a temporary register,
which is written back to $\matr{C}$ at the end of each iteration $m\in M$. The
multiplication of elements of $\matr{A}$ and $\matr{B}$ can be pipelined, but
the addition on \coderef{acc} requires the result of the addition in the
previous iteration of the loop. This is a loop-carried dependency, and results
in an initiation interval of $L_+$, where $L_+$ is the latency of a
$\SI{64}{\bit}$ floating point  addition (for integers
$L_{+,\text{\texttt{int}}}{=}\SI{1}{\cycle}$, and the loop can be pipelined
without further modifications). To avoid this, we can transpose the iteration
space, swapping the $K$-loop with the $M$-loop, with the following consequences:
\begin{itemize}[leftmargin=*]
  \item Rather than a single register, we now implement an accumulation buffer
  of depth $M$ and width $1$ (\coderef{acc_buffer}).
  \item The loop-carried dependency is resolved: each location is only updated
  every $M$ cycles (with $M{\geq}L_\text{acc}$ in
  \figureref{buffered_accumulation}).
  \item $\matr{A}$, $\matr{B}$, and $\matr{C}$ are all read in a contiguous
  fashion, achieving perfect spatial locality (we assume row-major memory
  layout. For column-major we would interchange the $K$-loop and $N$-loop).
  \item Each element of $\matr{A}$ is read exactly once. 
\end{itemize}
The modified code is shown in \listingref{gemm_transposed}. We leave the
accumulation buffer defined on \coderef{acc_buffer} uninitialized, and
implicitly reset it on \coderef{acc_reset}, avoiding $M$ extra cycles to
reset (this is a form of \emph{pipelined loop fusion}, covered in
\secref{pipelined_loop_fusion}). 

\vspace{-0.5em} 
\subsubsection{Tiled Accumulation Interleaving}
\vspace{-0.25em} 
\label{sec:partial_interleaving}
\label{sec:nested_accumulation_interleaving}

For accumulations done in a nested loop, it can be sufficient to interleave
across a tile of an outer loop to resolve a loop-carried dependency, using a
limited size buffer to store intermediate results. This tile only needs to be of
size ${\geq}L_\text{acc}$, where $L_\text{acc}$ is the latency of the
accumulation operation.

This is shown in \listingref{gemm_tiled}, for the transposed matrix
multiplication example from \listingref{gemm_transposed}, where the accumulation
array has been reduced to tiles of size $T$ (which should be
${\geq}L_\text{acc}$, see \figureref{buffered_accumulation}), by adding an
additional inner loop over the tile, and cutting the outer loop by a factor of
$B$.

\vspace{-0.5em} 
\subsubsection{Single-Loop Accumulation Interleaving}
\vspace{-0.25em} 

If no outer loop is present, we have to perform the accumulation in two separate
stages, at the cost of extra resources. For the first stage, we perform a
transformation similar to the nested accumulation interleaving, but strip-mine
the inner (and only) loop into blocks of size $K \geq L_\text{acc}$,
accumulating partial results into a buffer of size $K$. Once all incoming values
have been accumulated into the partial result buffers, the second
phase collapses the partial results into the final output. This is shown in
\listingref{two_stage_accumulation} for $K{=}16$.

\begin{listing}[h]
  \begin{minipage}{.57\columnwidth}
    \begin{minted}{C++}
double Acc(double arr[], int N) {
  double t[16];
  #pragma PIPELINE
  for (int i = 0; i < N; ++i) { // P0
    auto prev = (i < 16) ? 0 : t[i%16];
    t[i%16] = prev + arr[i]; }
  double res = 0;
  for (int i = 0; i < 16; ++i)  // P1
    res += t[i]; // Not pipelined
  return res; }
    \end{minted}
  \end{minipage}\hfill%
  \begin{minipage}{.41\columnwidth}
    \includegraphics[width=\columnwidth]{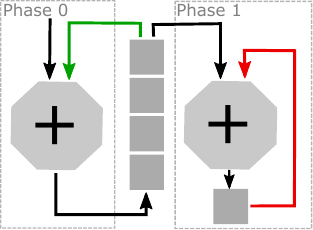}
  \end{minipage}
  \vspace{-0.5em}
  \caption{Two stages required for single loop accumulation.}
  \label{lst:two_stage_accumulation}
\end{listing}

\noindent Optionally, the two stages can be implemented to run in a
coarse-grained pipelined fashion, such that the first stage begins computing new
partial results while the second stage is collapsing the previous results (by
exploiting dataflow between modules, see \secref{dataflow}).

\vspace{-0.5em} 
\subsubsection{Batched Accumulation Interleaving}
\vspace{-0.25em} 
\label{sec:cross_input_pipelining}
\label{sec:batched_input_pipelining}

For algorithms with loop-carried dependencies that cannot be solved by either
method above (e.g., due to a non-commutative accumulation operator), we can
still pipeline the design by processing \emph{batches} of inputs, introducing an
additional loop nested in the accumulation loop. 
This procedure is similar to \secref{nested_accumulation_interleaving}, but only
applies to programs where it is relevant to compute the accumulation for
multiple data streams, and requires altering the interface and data movement of
the program to interleave inputs in batches.

The code in \listingref{cross_input_before} shows an iterative solver code with
an inherent loop-carried dependency on \texttt{state}, with a minimum initiation
interval corresponding to the latency $L_\text{Step}$ of the (inlined) function
\texttt{Step}. There are no loops to interchange, and we cannot change the order
of loop iterations. While there is no way to improve the latency of producing a
single result, we can improve the overall throughput by a factor of
$L_\text{Step}$ by pipelining across $N{\geq}L_\text{Step}$ different inputs
(e.g., overlap solving for different starting conditions). We effectively inject
another loop over inputs, then perform transposition or tiled accumulation
interleaving with this loop. The result of this transformation is shown in
\listingref{cross_input_after}, for a variable number of interleaved inputs
\texttt{N}.

\begin{listing}[h]
  \begin{sublisting}[b]{\columnwidth}
    \begin{minted}{C++}
Vec<double> IterSolver(Vec<double> state, int T) {
  #pragma PIPELINE // Will fail to pipeline with I=1 
  for (int t = 0; t < T; ++t)
    state = Step(state);
  return state; }
    \end{minted}
    \vspace{-0.75em}
    \caption{Solver executed for $T$ steps with a loop-carried dependency on
    \texttt{state}.}
    \label{lst:cross_input_before}
    \vspace{0.5em}
  \end{sublisting}\hfill
  \begin{sublisting}[b]{\columnwidth}
    \begin{minipage}{.63\columnwidth}
      \begin{minted}{C++}
template <int N>
void MultiSolver(Vec<double> *in,
                 Vec<double> *out, int T) {
  Vec<double> b[N]; // Partial results 
  for (int t = 0; t < T; ++t)
    #pragma PIPELINE
    for (int i = 0; i < N; ++i) {
      auto read = (t == 0) ? in[i] : b[i]; 
      auto next = Step(read);
      if (t < T-1) b[i] = next; 
      else out[i] = next; }} // Write out 
      \end{minted}
    \end{minipage}\hfill%
    \begin{minipage}{.37\columnwidth}
      \flushright
      \includegraphics[width=\textwidth]{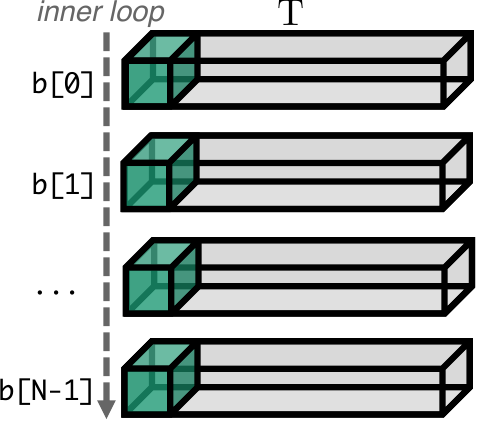}
    \end{minipage}
    \vspace{-0.5em}
    \caption{Pipeline across $N{\geq}L_\text{step}$ inputs to achieve
    $I{=}\SI{1}{\cycle}$.}
    \label{lst:cross_input_after}
    \vspace{0.5em}
  \end{sublisting}
  \vspace{-2em}
  \caption{Pipeline across multiple inputs to avoid loop-carried dependency.}
  \label{lst:cross_input_pipelining}
\end{listing}

\vspace{-1.75em} 
\subsection{Delay Buffering}
\vspace{-0.25em} 
\label{sec:delay_buffering}
\label{sec:shift_registers}
\label{sec:fifo_buffering}

\begin{listing}[b]
  \vspace{-1em} 
  \begin{sublisting}[b]{\columnwidth}
    \begin{minipage}{\columnwidth}
      \begin{minted}[escapeinside=||]{C++}
float north_buffer[M];  // Line |\label{code:north_buffer}|
float center_buffer[M]; // buffers|\label{code:center_buffer}|
float west, center; // Registers 
for (int i = 0; i < N; ++i) {
  #pragma PIPELINE
  for (int j = 0; j < M; ++j) {
    auto south = memory[i][j]; // Single memory read |\label{code:read_south}|
    auto north = north_buffer[j];         // Read line buffers |\label{code:pop_north}|
    auto east = center_buffer[(j + 1)%M]; // (with wrap around) |\label{code:pop_center}|
    if (i > 1 && j > 0 && j < M - 1) // Assume padding of 1
      result[i - 1][j] = 0.25*(north + west + south + east);
    north_buffer[j] = center; // Update both |\label{code:push_buffers_0}|
    center_buffer[j] = south; // line buffers |\label{code:push_buffers_1}|
    west = center; center = east; } } // Propagate registers |\label{code:shift_registers}|
      \end{minted}
    \end{minipage}
    \vspace{-0.5em}
    \caption{Delay buffering using cyclically indexed line buffers.}
    \label{lst:stencil_streams}
    \vspace{0.5em}
  \end{sublisting}\hfill
  \begin{sublisting}[b]{\columnwidth}
    \begin{minipage}{\columnwidth}
      \begin{minted}[escapeinside=||]{C++}
float sr[2*M + 1]; // Shift register buffer
for (int i = 0; i < N; ++i) {
  #pragma PIPELINE
  for (int j = 0; j < M; ++j) {
    #pragma UNROLL 
    for (int k = 0; k < 2*M; ++k) |\label{code:shift_shift_register_0}|
      sr[k] = sr[k + 1]; // Shift the array left |\label{code:shift_shift_register_1}|
    sr[2*M] = memory[i][j]; // Append to the front 
    if (i > 1 && j > 0 && j < M - 1) // Initialize/drain
      result[i-1][j] = 0.25*(sr[0] + sr[M-1] + sr[M+1] + sr[2*M]); } } |\label{code:constant_indices}|
      \end{minted}
    \end{minipage}
    \vspace{-0.5em}
    \caption{Delay buffering using an Intel-style shift register.}
    \label{lst:stencil_sliding}
  \end{sublisting}
  \vspace{-1.75em}
  \caption{Two ways of implementing delay buffering on an $N{\times}M$ grid.}
  \label{lst:stencil}
\end{listing}

\begin{listing*}
  \begin{sublisting}[b]{.35\textwidth}
    \begin{minted}[escapeinside=||, stripnl=false]{C++}
// Pipelined loops executed sequentially
for (int i = 0; i < N0; ++i) Foo(i, /*...*/); 
for (int i = 0; i < N1; ++i) Bar(i, /*...*/); 
    \end{minted}
    \centering
    \includegraphics[height=3.5em]{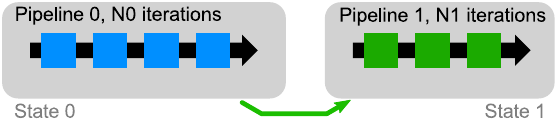}\\
    \vspace{-0.5em}
    \caption{$(L_0 + I_0 (N_0{-}1)) + (L_1 + I_1 (N_1{-}1))$ cycles.}
    \label{lst:fusion_initial}
  \end{sublisting}\hfill%
  \begin{sublisting}[b]{.29\textwidth}
    \begin{minted}[escapeinside=||]{C++}
for (int i = 0; i < N0+N1; ++i) {
  if (i < N0) Foo(i,      /*...*/);
         else Bar(i - N0, /*...*/); } 
    \end{minted}
    \centering
    \includegraphics[height=3.5em]{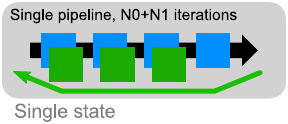}\\
    \vspace{-0.5em}
    \caption{$L_2 + I (N_0 + N_1{-}1)$ cycles.}
    \label{lst:fusion_sequential}
  \end{sublisting}\hfill%
  \begin{sublisting}[b]{.35\textwidth}
    \begin{minted}[escapeinside=||]{C++}
for (int i = 0; i < max(N0, N1); ++i) {
  if (i < N0) Foo(i, /*...*/);   // Omit ifs 
  if (i < N1) Bar(i, /*...*/); } // for N0==N1 
    \end{minted}
    \centering
    \includegraphics[height=3.5em]{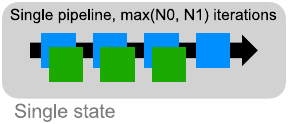}\\
    \vspace{-0.5em}
    \caption{$L_3 + I \cdot (\text{max}(N_0, N_1){-}1)$ cycles.}
    \label{lst:fusion_parallel}
  \end{sublisting}
  \vspace{-0.5em}
  \caption{Two subsequent pipelined loops fused sequentially
  (\listingref{fusion_sequential}) or concurrently
  (\listingref{fusion_parallel}).
  Assume that all loops are pipelined (pragmas omitted for brevity).}
  \label{lst:pipeline_fusion}
  \label{lst:pipelined_loop_fusion}
  \label{lst:fusion}
\end{listing*}

\noindent When iterating over regular domains in a pipelined fashion, it is
often sufficient to express buffering using delay buffers, expressed either with
cyclically indexed arrays, or with constant offset delay buffers, also known
from the Intel ecosystem as \emph{shift registers}.  These buffers are only
accessed in a FIFO manner, with the additional constraint that elements are only
be popped once they have \emph{fully} traversed the depth of the buffer (or when
they pass compile-time fixed access points, called ``taps'', in Intel OpenCL).
Despite the ``shift register'' name, these buffers do not need to be implemented
in registers, and are frequently implemented in on-chip RAM when large capacity
is needed, where values are not \emph{physically} shifted. 

A common set of applications that adhere to the delay buffer pattern are stencil
applications such as partial differential equation solvers~\cite{stencil_pde,
stencil_electrodynamics, stencil_cfd}, image processing
pipelines~\cite{darkroom, halide}, and convolutions in deep neural
networks~\cite{dnn_performance_survey, fpga_dnn_survey, binary_dnn,
hls_dnn_0, finn_r}, all of which are typically traversed using a
\emph{sliding window} buffer, implemented in terms of multiple delay buffers
(or, in Intel terminology, a shift register with multiple \emph{taps}). These
applications have been shown to be a good fit to spatial computing
architectures~\cite{reverse_time_migration, opencl_stencil_optimization_zohouri,
stencil_opencl_optimization_waidyasooriya, opencl_stencil_optimization_jia,
stencil_runtime_reconfig_0, stencil_runtime_reconfig_1, sliding_window_fpga}, as
delay buffering is cheap to implement in hardware, either as shift registers in
general purpose logic, or in RAM blocks.

\listingref{stencil} shows two ways of applying delay buffering to a stencil
code, namely a 4-point stencil in 2D, which updates each point on a 2D grid to
the average of its north, west, east, and south neighbors.  To achieve perfect
data reuse, we buffer every element read in sequential order from memory until
it has been used for the last time -- after two rows, when the same value has
been used as all four neighbors.

\begin{figure}[h]
  \centering
  \vspace{-0.25em}
  \begin{minipage}[b]{.99\columnwidth}
    \centering
    \includegraphics[height=4em]{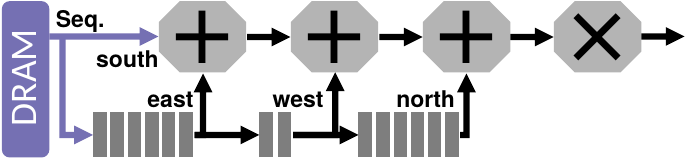}
    \vspace{-0.5em}
    \caption{A delay buffer for a 4-point stencil with three \emph{taps}.}
    \label{fig:jacobi_buffering}
  \end{minipage}\hfill%
  \centering
  \vspace{-0.25em}
\end{figure}

In \listingref{stencil_streams} we use cyclically indexed line buffers to
implement the delay buffering pattern, instantiated as arrays on
lines~\ref{code:north_buffer}-\ref{code:center_buffer}.  We only read the south
element from memory each iteration (\coderef{read_south}), which we store in the
center line buffer (\coderef{push_buffers_1}). This element is then reused after
$M$ cycles (i.e., ``delayed'' for $M$ cycles), when it is used as the east value
(\coderef{pop_center}), propagated to the north buffer
(\coderef{push_buffers_0}), shifted in registers for two cycles until it is used
as the west value (\coderef{shift_registers}), and reused for the last time
after $M$ cycles on \coderef{pop_north}. The resulting circuit is illustrated in
\figureref{jacobi_buffering}.

\listingref{stencil_sliding} demonstrates the shift register pattern used to
express the stencil buffering scheme, which is supported by the Intel OpenCL
toolflow. Rather than creating each individual delay buffer required to
propagate values, a single array is used, which is ``shifted'' every cycle using
unrolling
(lines~\ref{code:shift_shift_register_0}-\ref{code:shift_shift_register_1}). The
computation accesses elements of this array using \emph{constant indices only}
(\coderef{constant_indices}), relying on the tool to infer the partitioning into
individual buffers (akin to loop idiom
recognition~\cite{compiler_transformations}) that we did explicitly in
\listingref{stencil_streams}. The implicit nature of this pattern requires the
tool to specifically support it. For more detail on buffering stencil codes we
refer to other works on the subject~\cite{stencilflow,
stencil_opencl_optimization_waidyasooriya}.

\noindent Opportunities for delay buffering often arise naturally in pipelined
programs.  If we consider the transposed matrix multiplication code in
\listingref{gemm_transposed}, we notice that the read from \texttt{acc} on
\coderef{acc_reset} and the write on \coderef{acc_write} are both sequential,
and cyclical with a period of $M$ cycles. We could therefore also use the shift
register abstraction for this array.
The same is true for the accumulation code in \listingref{cross_input_after}. 



\vspace{-1.5em} 
\subsection{Random Access Buffering}
\vspace{-0.25em} 
\label{sec:ram_buffering}
\label{sec:random_buffering}

When a program unavoidably needs to perform random accesses, we can buffer data
in on-chip memory and perform random access to this fast memory instead of to
slow off-chip memory.
A random access buffer implemented with a general purpose replacement strategy
will emulate a CPU-style cache; but to benefit from targeting a spatial system,
it is usually more desirable to \emph{specialize} the buffering strategy to the
target application~\cite{bram_shuffling, light_propagation}.
This can enable off-chip memory accesses to be made contiguous by loading and
storing data in stages (i.e., tiles), then exclusively performing random
accesses to fast on-chip memory.

\listingref{histogram} outlines a histogram implementation that uses an on-chip
buffer (\coderef{histogram_buffer}) to perform fast random accesses reads and
writes (\coderef{fast_random_access}) to the bins computed from incoming data,
illustrated in \figureref{histogram_buffering}.
Note that the random access results in a loop-carried dependency on
\texttt{histogram}, as there is a potential for subsequent iterations to read
and write the same bin. This can be solved with one of the interleaving
techniques described in \secref{interleaving}, by maintaining multiple partial
result buffers. 

\begin{listing}[H]
  \begin{minipage}{.64\columnwidth}
    \begin{minted}[escapeinside=||]{C++}
unsigned hist[256] = {0}; // Array of bins|\label{code:histogram_buffer}|
#pragma PIPELINE // Will have II=2
for (int i = 0; i < N; ++i) {
  int bin = CalculateBin(memory[i]); 
  hist[bin] += 1; // Single cycle access|\label{code:fast_random_access}|
} // ...write result out to memory...
    \end{minted}
  \end{minipage}\hfill%
  \begin{minipage}{.35\columnwidth}
    \centering
    \includegraphics[height=5.5em]{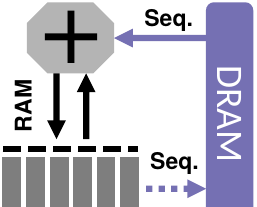}
  \end{minipage}
  \vspace{-0.5em}
  \caption{Random access to on-chip histogram buffer.}
  \label{lst:histogram}
  \label{fig:histogram_buffering}
\end{listing}

\vspace{-2em} 
\subsection{Pipelined Loop Fusion}
\vspace{-0.25em} 
\label{sec:fusion}
\label{sec:pipeline_fusion}
\label{sec:pipelined_loop_fusion}

When two pipelined loops appear sequentially, we can fuse them into a single
pipeline, while using loop guards to enforce any dependencies that might exist
between them. This can result in a significant reduction in runtime, at little
to no resource overhead. This transformation is closely related to
loop~fusion~\cite{dependence_graphs} from traditional software optimization.

For two consecutive loops with latencies/bounds/initiation intervals $\{L_0,
N_0, I_0\}$ and $\{L_1, N_1, I_1\}$ (\listingref{fusion_initial}), respectively,
the total runtime according to \equref{pipeline} is $(L_0 + I_0 (N_0{-}1)) +
(L_1 + I_1 (N_1{-}1))$. Depending on which condition(s) are met, we can
distinguish between three levels of pipelined loop fusion, with increasing
performance benefits: 
\begin{enumerate}[leftmargin=*]
  \item \emph{$I{=}I_0{=}I_1$} (true in most cases): Loops can be fused by
  summing the loop bounds, using loop guards to sequentialize them within the
  same pipeline (\listingref{fusion_sequential}). \label{enum:cond0}
  \item \label{enum:cond1} Condition~\ref{enum:cond0} is met, \textbf{and}
  \emph{only fine-grained or no dependencies} exist between the two loops: Loops
  can be fused by iterating to the maximum loop bound, and loop guards are
  placed as necessary to predicate each section (\listingref{fusion_parallel}).
  \item Conditions~\ref{enum:cond0}~and~\ref{enum:cond1} are met, \textbf{and}
  $N{=}N_0{=}N_1$ (same loop bounds): Loops bodies can be trivially fused
  (\listingref{fusion_parallel}, but with no loop guards necessary).
\end{enumerate}

\noindent An alternative way of performing pipeline fusion is to instantiate
each stage as a separate processing element, and stream fine-grained
dependencies between them (\secref{dataflow}).

\vspace{-0.5em} 
\subsection{Pipelined Loop Switching}
\label{sec:pipeline_switching}
\label{sec:loop_switching}
\label{sec:pipelined_loop_switching}

The benefits of pipelined loop fusion can be extended to coarse-grained control
flow by using \emph{loop switching} (as opposed to loop \emph{un}switching,
which is a common transformation~\cite{compiler_transformations} on load/store
architectures). 
Whereas instruction-based architectures attempt to only execute one branch of a
conditional jump (via branch prediction on out-of-order processors), a
conditional in a pipelined scenario will result in \emph{both} branches being
instantiated in hardware, regardless of whether/how often it is executed. The
transformation of coarse-grained control flow into fine-grained control flow is
implemented by the HLS tool by introducing \emph{predication} to the pipeline,
at no significant runtime penalty.

\listingref{pipelined_loop_switching} shows a simple example of how the
transformation fuses two pipelined loops in different branches into a single
loop switching pipeline. The transformation applies to \emph{any} pipelined code
in either branch, following the principles described for pipelined loop fusion
(\seclink{pipelined_loop_fusion} and \listingref{pipelined_loop_fusion}).

\begin{listing}[b]
  \begin{sublisting}[b]{.49\columnwidth}
    \centering
    \begin{minted}[escapeinside=||]{C++}
if (condition)
  #pragma HLS PIPELINE
  for (int i = 0; i < N0; ++i)
    y[i] = Foo(x[i]);
else
  #pragma HLS PIPELINE
  for (int i = 0; i < N1; ++i)
    y[i] = Bar(x[i]);
    \end{minted}
    \vspace{-0.5em}
    \caption{Coarse-grained control flow.}
    \label{lst:loop_switching_before}
  \end{sublisting}\hfill
  \begin{sublisting}[b]{.49\columnwidth}
    \centering
    \begin{minted}[escapeinside=||]{C++}
auto N = condition ? N0 : N1; |\label{code:loop_switching_bound}|
#pragma HLS PIPELINE
for (int i = 0; i < N; ++i) {
  if (condition)
    y[i] = Foo(x[i]);
  else
    y[i] = Bar(x[i]);
}
    \end{minted}
    \vspace{-0.5em}
    \caption{Control flow absorbed into pipeline.}
    \label{lst:loop_switching_after}
  \end{sublisting}
  \vspace{-0.5em}
  \caption{Pipelined loop switching absorbs coarse-grained control flow.}
  \label{lst:pipelined_loop_switching}
\end{listing}

The implications of pipelined loop switching are more subtle than the pure
fusion examples in \listingref{fusion}, as the total number of loop iterations
is not affected (assuming the fused loop bound is set according to the
condition, see line~\ref{code:loop_switching_bound} in
\listingref{loop_switching_after}).  There \emph{can} be a (tool-dependent)
benefit from saving overhead logic by only implementing the orchestration and
interfaces of a single pipeline, at the (typically minor) cost of the
corresponding predication logic.  More importantly, eliminating the
coarse-grained control can enable other transformations that significantly
benefit performance, such as fusion [\seclink{pipelined_loop_fusion}] with
adjacent pipelined loops, flattening nested loops [\seclink{flattening}], and
on-chip dataflow [\seclink{dataflow}]. 

\vspace{-0.5em} 
\subsection{Pipelined Loop Flattening/Coalescing}
\label{sec:flattening}
\label{sec:loop_flattening}

To minimize the number of cycles spent in filling/draining pipelines (where the
circuit is not streaming at full throughput), we can flatten nested loops to
move the fill/drain phases to the outermost loop, fusing/absorbing code that is
not in the innermost loop if necessary.

\listingref{loop_coalescing_before} shows a code with two nested loops, and
gives the total number of cycles required to execute the program.  The latency
of the drain phase of the inner loop and the latency of \texttt{Bar} outside
the inner loop must be \emph{paid at every iteration} of the outer loop.
If $N_0 {\gg} L_0$, the cycle count becomes just $L_1 + N_0 N_1$, but for
applications where $N_0$ is comparable to $L_0$, draining the inner pipeline
can significantly impact the runtime (even if $N_1$ is large).  By transforming
the code such that all loops are \emph{perfectly nested} (see
\listingref{loop_coalescing_after}), the HLS tool can effectively
\emph{coalesce} the loops into a single pipeline, where next iteration of the
\emph{outer} loop can be executed immediately after the previous finishes. 

\begin{listing}[h]
  \begin{sublisting}[b]{.47\columnwidth}
    \begin{minted}[escapeinside=||]{C++}
for (int i = 0; i < N1; ++i) {
  #pragma PIPELINE
  for (int j = 0; j < N0; ++i)
    Foo(i, j); 
  Bar(i); } |\label{code:drain_code}|
    \end{minted}
    \vspace{-0.5em}
    \caption{$L_1 + N_1 \cdot (L_0 + N_0{-}1)\;\si{\cycles}$.}
    \label{lst:loop_coalescing_before}
  \end{sublisting}\hfill%
  \begin{sublisting}[b]{.53\columnwidth}
    \begin{minted}[escapeinside=||]{C++}
for (int i = 0; i < N1; ++i) {
  #pragma PIPELINE
  for (int j = 0; j < N0; ++i)
    Foo(i, j); 
    if (j == N0 - 1) Bar(i); } |\label{code:loop_guard}|
    \end{minted}
    \vspace{-0.5em}
    \caption{$L_2 + N_0 N_1{-}1\;\si{\cycles}$.}
    \label{lst:loop_coalescing_after}
  \end{sublisting}
  \begin{minipage}{.49\columnwidth}
    \includegraphics[width=\textwidth]{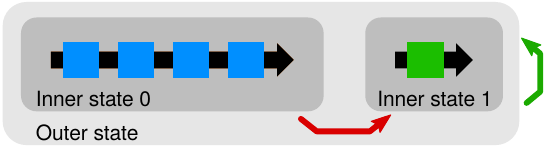}
  \end{minipage}\hfill%
  \begin{minipage}{.49\columnwidth}
    \includegraphics[width=\textwidth]{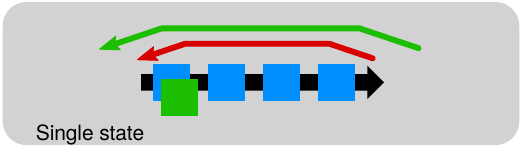}
  \end{minipage}
  \vspace{-0.5em}
  \caption{Before and after coalescing loop nest to avoid inner pipeline drains.}
  \label{lst:loop_coalescing}
\end{listing}

\noindent To perform the transformation in \listingref{loop_coalescing}, we had
to absorb \texttt{Bar} into the inner loop, adding a loop guard
(\coderef{loop_guard} in \listingref{loop_coalescing_after}), analogous to
pipelined loop fusion (\seclink{fusion}), where the second pipelined ``loop''
consists of a single iteration. This contrasts the loop peeling transformation,
which is used by CPU compilers to regularize loops to avoid branch
mispredictions and increasing amenability to vectorization. While loop peeling
can also be beneficial in hardware, e.g., to avoid deep conditional logic in a
pipeline, small inner loops can see a significant performance improvement by
eliminating the draining phase.

\vspace{-0.75em} 
\subsection{Inlining}
\label{sec:inlining}

In order to successfully pipeline a scope, all function calls within the code
section must be pipelineable.
This typically requires ``inlining'' functions into each call site,
creating dedicated hardware for each invocation, resulting in additional
resources consumed for every additional callsite after the first.
This replication is done automatically by HLS compilers on demand, but an
additional \texttt{inline} pragma can be specified to directly ``paste'' the
function body into the callsite during preprocessing, removing the function
boundary during optimization and scheduling.

\vspace{-1em} 
\section{Scalability Transformations}
\label{sec:scalability}

Parallelism in HLS revolves around the \emph{folding} of loops, achieved through
\emph{unrolling}.
In \secref{interleaving} we used strip-mining and reordering to avoid
loop-carried dependencies by changing the \emph{schedule} of computations in the
pipelined loop nest. 
In this section, we similarly strip-mine and reorder loops, but with additional
unrolling of the strip-mined chunks.
Pipelined loops constitute the \emph{iteration space}; the size of which
determines the number of cycles it takes to execute the program. Unrolled loops,
in a pipelined program, correspond to the degree of \emph{parallelism} in the
architecture, as every expression in an unrolled statement is required to exist
as hardware.
Parallelizing a code thus means turning sequential/pipelined loops fully or
partially into parallel/unrolled loops. This corresponds to \emph{folding} the
sequential iteration space, as the number of cycles taken to execute the program
are effectively reduced by the inverse of the unrolling factor.

\begin{figure}[h]
  \begin{subfigure}[b]{.13\columnwidth}
    \centering
    \includegraphics[height=2.2em]{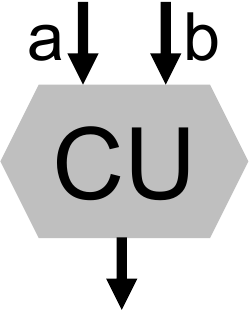}
    \caption{Before.}
    \label{fig:before_vectorization}
  \end{subfigure}\hfill%
  \begin{subfigure}[b]{.29\columnwidth}
    \centering
    \includegraphics[height=2.2em]{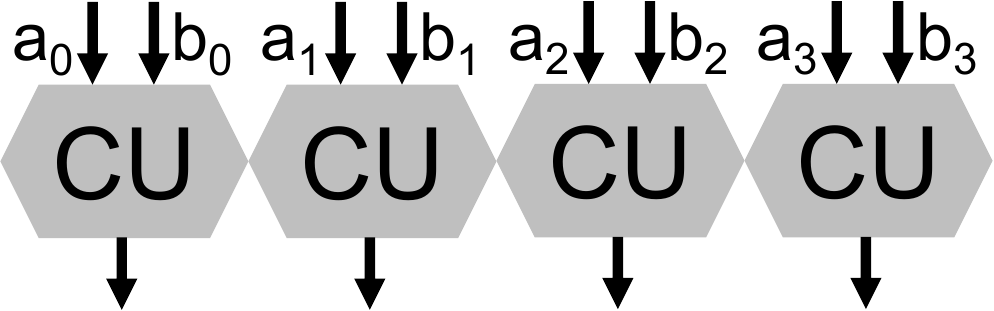}
    \caption{Horizontal unroll.}
    \label{fig:after_vectorization}
  \end{subfigure}\hfill%
  \begin{subfigure}[b]{.29\columnwidth}
    \centering
    \includegraphics[height=2.2em]{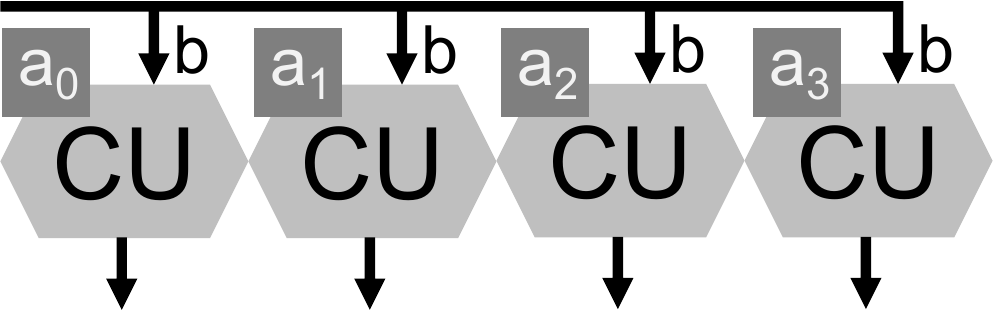}
    \caption{Vertical unroll.}
    \label{fig:after_replication}
  \end{subfigure}\hfill%
  \begin{subfigure}[b]{.29\columnwidth}
    \centering
    \includegraphics[height=2.2em]{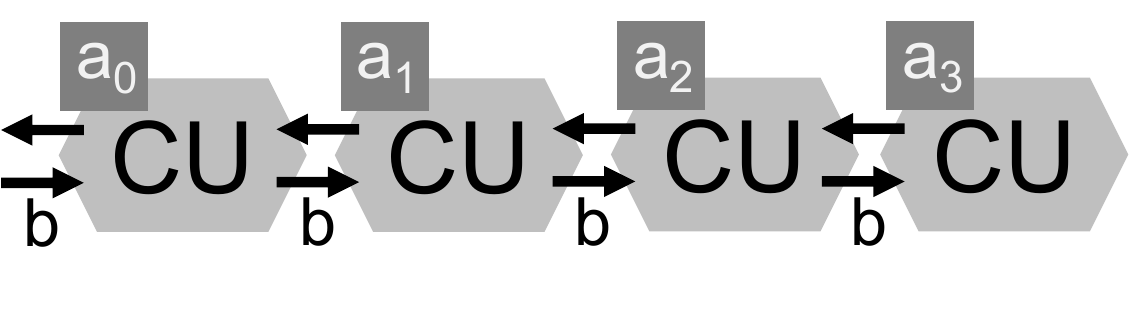}
    \caption{Dataflow.}
    \label{fig:after_dataflow}
  \end{subfigure}
  \caption{Horizontal unrolling, vertical unrolling, and dataflow, as means to
  increase parallelism. Rectangles represent buffer space, such as registers or
  on-chip RAM. \textbf{Horizontal:} four independent inputs processed in
  parallel. \textbf{Vertical:} one input is combined with multiple buffered
  values. \textbf{Dataflow:} similar to vertical, but input or partial results
  are streamed through a pipeline rather than broadcast.}
  \label{fig:parallelism}
\end{figure}

\vspace{-1.5em} 
\subsection{Horizontal Unrolling (Vectorization)}
\label{sec:vectorization}
\label{sec:horizontal_unrolling}

We implement vectorization-style parallelism with HLS by ``horizontally''
unrolling loops in pipelined sections, or by introducing vector types, folding
the sequential iteration space accordingly.
This is the most straightforward way of adding parallelism, as it can often be
applied directly to an inner loop without further reordering or drastic changes
to the nested loop structure.
Vectorization is more powerful in HLS than SIMD operations on load/store
architectures, as the unrolled compute units are not required to be homogeneous,
and the number of units are not constrained to fixed sizes.
Horizontal unrolling increases bandwidth utilization by explicitly exploiting
spatial locality, allowing more efficient accesses to off-chip memory such as
DRAM.

\listingref{simd_unrolling} shows two functionally equivalent ways of
vectorizing a loop over $N$ elements by a horizontal unrolling factor of $W$.
\listingref{simd_unrolling_two_loops} strip-mines a loop into chunks of $W$ and
unrolls the inner loop fully, while \listingref{simd_unrolling_partial_unroll}
uses partial unrolling by specifying an unroll factor in the pragma. As a third
option, explicit vector types can be used, such as those built into OpenCL
(e.g., \texttt{float4} or \texttt{int16}), or custom vector
classes~\cite{hlslib}. These provide less flexibility, but are more concise and
are sufficient for most applications.

\begin{listing}[h]
  \begin{sublisting}[b]{.57\columnwidth}
    \begin{minted}{C++}
for (int i = 0; i < N / W; ++i)
  #pragma UNROLL // Fully unroll inner
  for (int w = 0; w < W; ++w) // loop
    C[i*W + w] = A[i*W + w]*B[i*W + w];
    \end{minted}
    \vspace{-0.5em}
    \caption{Using strip-mining.}
    \label{lst:simd_unrolling_two_loops}
  \end{sublisting}\hfill%
  \begin{sublisting}[b]{.42\columnwidth}
    \begin{minted}{C++}
// Unroll outer loop by W 
#pragma UNROLL W 
for (int i = 0; i < N; ++i)
  C[i] = A[i] * B[i];
    \end{minted}
    \vspace{-0.5em}
    \caption{Using partial unrolling.}
    \label{lst:simd_unrolling_partial_unroll}
  \end{sublisting}
    \vspace{-0.5em}
  \caption{Two variants of vectorization by factor $W$ using loop unrolling.}
  \label{lst:simd_unrolling}
\end{listing}

\noindent In practice, the unrolling factor $W\;[\si{\operand\per\cycle}]$ is
constrained by the bandwidth $B\;[\si{\byte\per\second}]$ available to the
compute logic (e.g., from off-chip memory), according to $W_\text{max} =
\floor*{\frac{B}{f S}}$,
%
%
where $f\;[\si{\cycle\per\second}]$ is the clock frequency of the unrolled
logic, and $S\;[\si{\byte\per\operand}]$ is the operand size in bytes.
Horizontal unrolling is usually not sufficient to achieve high logic utilization
on large chips, where the available memory bandwidth is low compared to the
available amount of compute logic.  Furthermore, because the energy cost of I/O
is orders of magnitude higher than moving data on the chip, it is desirable to
exploit on-chip memory and pipeline parallelism instead (this follows in
Sec.~\ref{sec:replication}~and~\ref{sec:streaming}). 

\vspace{-1em} 
\subsection{Vertical Unrolling}
\label{sec:replication}
\label{sec:vertical_unrolling}

We can achieve scalable parallelism in HLS without relying on external memory
bandwidth by exploiting data reuse, distributing input elements to multiple
computational units replicated ``vertically'' through
unrolling~\cite{opencl_smith_waterman_0, opencl_stencil_optimization_zohouri,
matrix_multiplication_definelicht}. \emph{This is the most potent source of
parallelism on hardware architectures}, as it can conceptually scale
indefinitely with available silicon when enough reuse is possible.  Viewed from
the paradigm of cached architectures, the opportunity for this transformation
arises from temporal locality in loops.
Vertical unrolling draws on bandwidth from on-chip fast memory by storing more
elements temporally, combining them with new data streamed in from external
memory to increase parallelism, allowing more computational units to run in
parallel at the expense of buffer space. In comparison, horizontal unrolling
requires us to widen the data path that passes through the processing elements
(compare \figureref{after_vectorization}~and~\ref{fig:after_replication}). 

When attempting to parallelize a new algorithm, identifying a source of temporal
parallelism to feed vertical unrolling is essential to whether the design will
scale. Programmers should consider this carefully before designing the hardware
architecture. From a reference software code, the programmer can identify
scenarios where reuse occurs, then extract and \emph{explicitly express} the
temporal access pattern in hardware, using a delay buffering 
[\seclink{fifo_buffering}] or random-access [\seclink{ram_buffering}] buffering
scheme. Then, if additional reuse is possible, vertically unroll the circuit
to scale up performance.

\begin{listing}[b]
  \centering
  \begin{mintedBlock}
    \begin{cpart}{}
for (int n = 0; n < N / P; ++n) { // Folded by unrolling factor P |\label{code:gemm_row_blocks}|
  for (int m = 0; m < M / T; ++m) { // Tiling 
    double acc[T][P]; // Is now 2D
    // ...initialize acc from C...
    for (int k = 0; k < K; ++k) {
      double a_buffer[P]; // Buffer multiple elements to combine with
      #pragma PIPELINE    // incoming values of B in parallel
    \end{cpart}
    \begin{cpart}{bgcolor=\mintedalt}
      for (int p = 0; p < P; ++p) |\label{code:buffer_a_begin}|
        a_buffer[p] = A[n*P + p][k]; |\label{code:buffer_a_end}|
      \end{cpart}
      \begin{cpart}{}
      #pragma PIPELINE
      for (int t = 0; t < T; ++t) // Stream tile of B
      \end{cpart}
      \begin{cpart}{bgcolor=\mintedalt}
        #pragma UNROLL |\label{code:replication_start}|
        for (int p = 0; p < P; ++p) // P-fold vertical unrolling
          acc[t][p] += a_buffer[p] * B[k][m*T+t]; |\label{code:replication_end}|
      \end{cpart}
      \begin{cpart}{}
    } /* ...write back 2D tile of C... */ } }
    \end{cpart}
  \end{mintedBlock}
  \vspace{-1em}
  \caption{$P$-fold vertical unrolling of matrix multiplication.}
  \label{lst:gemm_replication}
\end{listing}

As an example, we return to the matrix multiplication code from
\listingref{gemm_tiled}.  In \secref{partial_interleaving}, we saw that
strip-mining and reordering loops allowed us to move reads from matrix
$\matr{A}$ out of the inner loop, re-using the loaded value across $T$
different entries of matrix $\matr{B}$ streamed in while keeping the element of
$\matr{A}$ in a register. Since every loaded value of $\matr{B}$
\emph{eventually} needs to be combined with all $N$ rows of $\matr{A}$, we
realize that we can perform more computations in parallel by keeping
\emph{multiple} values of $\matr{A}$ in local registers. The result of this
transformation is shown in \listingref{gemm_replication}. By buffering $P$
elements (where $P$ was $1$ in \listingref{gemm_tiled}) of $\matr{A}$ prior to
streaming in the tile of $\matr{B}$-matrix
(lines~\ref{code:buffer_a_begin}-\ref{code:buffer_a_end}), we can \emph{fold}
the outer loop over rows by a factor of $P$, using unrolling to multiply
parallelism (as well as buffer space required for the partial sums) by a factor
of $P$ (lines~\ref{code:replication_start}-\ref{code:replication_end}). 

\vspace{-1em} 
\subsection{Dataflow}
\label{sec:dataflow}
\label{sec:streaming}
\label{sec:streaming_dataflow}

For complex codes it is common to partition functionality into multiple modules,
or \emph{processing elements} (PEs), streaming data between them through
explicit interfaces. In contrast to conventional pipelining, PEs arranged in a
dataflow architecture are scheduled separately when synthesized by the HLS tool.
There are multiple benefits to this:
\begin{itemize}[leftmargin=*]
  \item \emph{Different functionality runs at different schedules}. For example,
  \emph{issuing} memory requests, \emph{servicing} memory requests, and
  \emph{receiving} requested memory can all require different pipelines, state
  machines, and even clock rates. 
  \item Smaller components are more \emph{modular}, making them easier to reuse,
  debug and verify. 
  \item The effort required by the HLS tool to schedule code sections increases
  dramatically with the number of operations that need to be considered for the
  dependency and pipelining analysis.  Scheduling logic in smaller chunks is
  thus beneficial for compilation time. 
  \item Large \emph{fan-out}/\emph{fan-in} is challenging to route on real
  hardware, (i.e., $1$-to-$N$ or $N$-to-$1$ connections for large $N$).  This is
  mitigated by partitioning components into smaller parts and adding more
  pipeline stages. 
	\item The fan-in and fan-out of control signals (i.e., stall, reset)
\emph{within} each module is reduced, reducing the risk of these signals
constraining the maximum achievable frequency.
\end{itemize}
To move data between PEs, communication channels with a handshake mechanism are
used. These channels double as synchronization points, as they imply a
consensus on the program state. In practice, channels are always FIFO
interfaces, and support standard queue operations \texttt{Push}, \texttt{Pop},
and sometimes \texttt{Empty}, \texttt{Full}, and \texttt{Size} operations. They
occupy the same register or block memory resources as other buffers
(\secref{fifo_buffering}/\secref{random_buffering}).

The mapping from source code to PEs differs between HLS tools, but is
manifested when functions are connected using channels. In the following
example, we will use the syntax from Xilinx~Vivado~HLS to instantiate PEs,
where each non-inlined function correspond to a PE, and these are connected by
channels that are passed as arguments to the functions from a top-level entry
function. Note that this \textbf{functionally diverges from C++ semantics}
without additional abstraction~\cite{hlslib}, as each function in the dataflow
scope is executed in parallel in hardware, rather than in the sequence
specified in the imperative code. In Intel~OpenCL, dataflow semantics are
instead expressed with multiple \texttt{kernel} functions each defining a PE,
which are connected by global channel objects prefixed with the
\texttt{channel} keyword.

\begin{listing}[b]
  \begin{sublisting}[b]{\columnwidth}
    \begin{minted}[escapeinside=||]{C++}
void PE(FIFO<float> &in, FIFO<float> &out, int T) {
  // ..initialization...
  for (int t = 0; t < T / P; ++t) // Fold timesteps T by factor P |$\label{code:stencil_folded}$|
    #pragma PIPELINE
    for (/* loops over spatial dimensions */) {
      auto south = in.Pop(); // Value for t-1 from previous PE 
      // ...load values from delay buffers...
      auto next = 0.25*(north + west + east + south); 
      out.Push(next); }} // Value for t sent to PE computing t+1 
    \end{minted}
    \vspace{-0.5em}
    \caption{Processing element for a single timestep. Will be replicated $P$
    times.}
    \vspace{0.5em}
    \label{lst:stencil_processing_element}
    \label{lst:stencil_pe}
  \end{sublisting}\hfill
  \begin{sublisting}[b]{\columnwidth}
    \begin{minted}{C++}
#pragma DATAFLOW // Schedule nested functions as parallel modules 
void SystolicStencil(const float in[], float out[], int T) {
  FIFO<float> pipes[P + 1]; // Assume P is given at compile time
  ReadMemory(in, pipes[0]); // Head
  #pragma UNROLL // Replicate PEs
  for (int p = 0; p < P; ++p)
    PE(pipe[p], pipe[p + 1], T); // Forms a chain
  WriteMemory(pipes[P], out); } // Tail
    \end{minted}
    \vspace{-0.5em}
    \caption{Instantiate and connect $P$ consecutive and parallel PEs.}
    \label{lst:stencil_toplevel}
  \end{sublisting}
  \vspace{-1.5em}
  \caption{Dataflow between replicated PEs to compute $P$ timesteps in
  parallel.}
  \label{lst:stencil_dataflow}
\end{listing}

To see how streaming can be an important tool to express scalable hardware, we
apply it in conjunction with vertical unrolling (\secref{replication}) to
implement an iterative version of the stencil example from \listingref{stencil}.
Unlike the matrix multiplication code, the stencil code has no scalable source
of parallelism in the spatial dimension. Instead, we can achieve reuse by
folding the outer time-loop to treat $P$ consecutive timesteps in a pipeline
parallel fashion, each computed by a distinct PE, connected in a chain via
channels~\cite{reverse_time_migration, multi_fpga_stencil,
opencl_stencil_optimization_zohouri}.
%
We replace the memory interfaces to the PE with channels, such that the memory
read and write become \texttt{Pop} and \texttt{Push} operations, respectively.
The resulting code is shown in \listingref{stencil_processing_element}. We then
vertically unroll to generate $P$ instances of the PE (shown in
\listingref{stencil_toplevel}), effectively increasing the throughput of the
kernel by a factor of $P$, and consequently reducing the runtime by folding the
outermost loop by a factor of $P$ (\coderef{stencil_folded} in
\listingref{stencil_pe}).  Such architectures are sometimes referred to as
\emph{systolic arrays}~\cite{systolic_arrays, stencil_dsl}.

For architectures/HLS tools where large fan-out is an issue for compilation or
routing, an already replicated design can be transformed to a dataflow
architecture. For example, in the matrix multiplication example in
\listingref{gemm_replication}, we can move the $P$-fold unroll out of the inner
loop, and replicate the entire PE instead, replacing reads and writes with
channel accesses~\cite{matrix_multiplication_definelicht}.  $\matr{B}$ is then
streamed into the first PE, and passed downstream every cycle. $\matr{A}$ and
$\matr{C}$ should no longer be accessed by every PE, but rather be handed
downstream similar to $\matr{B}$, requiring a careful implementation of the
start and drain phases, where the behavior of each PE will vary slightly
according to its depth in the sequence. 

\vspace{-0.75em} 
\subsection{Tiling}
\vspace{-0.25em} 
\label{sec:tiling}

Loop tiling in HLS is commonly used to fold large problem sizes into manageable
chunks that fit into fast on-chip memory, in an already pipelined
program~\cite{opencl_stencil_optimization_zohouri}.
Rather than making the program faster, this lets the already fast architecture
support arbitrarily large problem sizes. This is in contrast to loop tiling on
CPU and GPU, where tiling is used to increase performance. Common to both
paradigms is that they fundamentally aim to meet fast memory constraints.
As with horizontal and vertical unrolling, tiling relies on strip-mining loops
to alter the iteration space.

Tiling was already shown in \secref{partial_interleaving}, when the
accumulation buffer in \listingref{gemm_transposed} was reduced to a tile
buffer in \listingref{gemm_tiled}, such that the required buffer space used for
partial results became a constant, rather than being dependent on the input
size.  This transformation is also relevant to the stencil codes in
\listingref{stencil}, where it can be used similarly to restrict the size of
the line buffers or shift register, so they a no longer proportional to the
problem size. 

\vspace{-0.75em} 
\section{Memory Access Transformations}
\vspace{-0.25em} 
\label{sec:memory_access_enhancing_transformations}
\label{sec:memory_transformations}

When an HLS design has been pipelined, scheduled, and unrolled as desired, the
memory access pattern has been established. In the following, we describe
transformations that optimize the efficiency of off-chip memory accesses in the
HLS code. For memory bound codes in particular, this is critical for
performance after the design has been pipelined.

\vspace{-0.75em} 
\subsection{Memory Access Extraction}
\label{sec:extraction}
\label{sec:memory_extraction}
\label{sec:memory_access_extraction}

By extracting accesses to external memory from the computational logic, we
enable compute and memory accesses to be pipelined and optimized separately.
Accessing the same interface multiple times within the same pipelined section is
a common cause for poor memory bandwidth utilization and increased initiation
interval due to interface contention, since the interface can only service a
single request per cycle. In the Intel OpenCL flow, memory extraction is done
automatically by the tool, but since this process must be conservative due to
limited information, it is often still beneficial to do the extraction
explicitly in the code~\cite{galerkin}.
In many cases, such as for independent reads, this is not an inherent memory
bandwidth or latency constraint, but arises from the tool scheduling iterations
according to program order. This can be relaxed when allowed by inter-iteration
dependencies (which can in many cases be determined automatically, e.g., using
polyhedral analysis~\cite{polly}).

\begin{listing}[b]
  \begin{sublisting}[b]{.99\columnwidth}
    \begin{minipage}{.65\columnwidth}
      \begin{minted}{C++}
void PE(const int A[N], int B[N/2]) {
  #pragma PIPELINE // Achieves I=2
  for (int i = 0; i < N/2; ++i)
    // Issues N/2 memory requests of size 1
    B[i] = A[i] + A[N/2 + i];
}
      \end{minted}
    \end{minipage}\hfill%
    \begin{minipage}{.34\columnwidth}
      \flushright
      \includegraphics[height=6em]{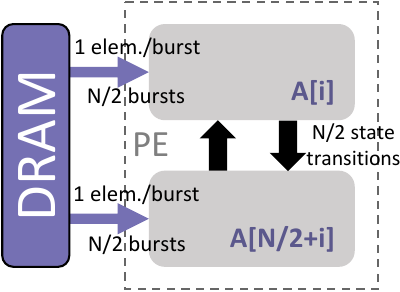}
    \end{minipage}
    \vspace{-0.5em}
    \caption{Multiple accesses to $A$ cause inefficient memory accesses.}
    \label{lst:before_memory_extraction}
  \end{sublisting}
  \begin{sublisting}[b]{.99\columnwidth}
    \begin{minipage}{.55\columnwidth}
      \begin{minted}[escapeinside=||]{C++}
void PE(FIFO<int> &A0, FIFO<int> &A1,
        int B[N/2]) {
  #pragma PIPELINE // Achieves I=1
  for (int i = 0; i < N/2; ++i)
    B[i] = A0.Pop() + A1.Pop()); |$\label{code:parallel_pop}$|
}
      \end{minted}
    \end{minipage}\hfill%
    \begin{minipage}{.44\columnwidth}
      \flushright
      \includegraphics[height=6em]{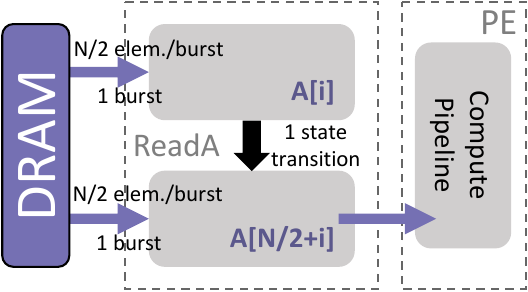}
    \end{minipage}
    \vspace{-0.5em}
    \caption{Move memory accesses out of computational code.}
    \label{lst:with_memory_extraction}
  \end{sublisting}
  \begin{sublisting}[b]{.99\columnwidth}
    \begin{minted}{C++}
void ReadA(const int A[N], FIFO<int> &A0, FIFO<int> &A1) {
  int buffer[N/2];
  #pragma PIPELINE
  for (int i = 0; i < N/2; ++i)
    buffer[i] = A[i]; // Issues 1 memory request of size N/2 
  #pragma PIPELINE
  for (int i = 0; i < N/2; ++i) {
    A0.Push(buffer[i]); // Sends to PE
    A1.Push(A[N/2 + i]); }} // Issues 1 memory request of size N/2 
    \end{minted}
    \vspace{-0.5em}
    \caption{Read $A$ in long bursts and stream them to the PE.}
    \label{lst:after_memory_extraction}
  \end{sublisting}
  \vspace{-0.5em}
  \caption{Separate memory accesses from computational logic.}
  \label{lst:memory_extraction}
\end{listing}

In \listingref{before_memory_extraction}, the same memory (i.e., hardware memory
interface) is accessed twice in the inner loop. In the worst case, the program 
will issue two $\SI{4}{\byte}$ memory requests every iteration, resulting in
poor memory performance, and preventing pipelining of the loop.
In software, this problem is typically mitigated by caches, always fetching at
least one cache line.
If we instead read the two sections of $A$ sequentially (or in larger chunks),
the HLS tool can infer two bursts accesses to $A$ of length $N/2$, shown in
\listingref{after_memory_extraction}. Since the schedules of memory and
computational modules are independent, \texttt{ReadA} can run ahead of
\texttt{PE}, ensuring that memory is always read at the maximum bandwidth of
the interface (\secref{memory_buffering} and \secref{striping} will cover how
to increase this bandwidth).  From the point of view of the computational PE,
both $A_0$ and $A_1$ are read in parallel, as shown on \coderef{parallel_pop}
in \listingref{with_memory_extraction}, hiding initialization time and
inconsistent memory producers in the synchronization implied by the data
streams.

An important use case of memory extraction appears in the stencil code in
\listingref{stencil_dataflow}, where it is necessary to separate the memory
accesses such that the PEs are agnostic of whether data is produced/consumed by
a neighboring PE or by a memory module.  
Memory access extraction is also useful for performing data layout
transformations in fast on-chip memory. For example, we can change the schedule
of reads from $\matr{A}$ in \listingref{gemm_replication} to a more efficient
scheme by buffering values in on-chip memory, while streaming them to the kernel
according to the original schedule.

\vspace{-0.75em} 
\subsection{Memory Buffering}
\vspace{-0.25em} 
\label{sec:memory_buffering}
\label{sec:memory_oversubscription}
\label{sec:oversubscription}

When dealing with memory interfaces with an inconsistent data rate, such as
DRAM, it can be beneficial to request and buffer accesses earlier and/or at a
more aggressive pace than what is consumed or produced by the computational
elements.  For memory reads, this can be done by reading ahead of the kernel
into a deep buffer instantiated between memory and computations, by either 1)
accessing wider vectors from memory than required by the kernel, narrowing or
widening data paths (aka. ``gearboxing'') when piping to or from computational
elements, respectively, or 2) increasing the clock rate of modules accessing
memory with respect to the computational elements.

The memory access function \listingref{after_memory_extraction} allows long
bursts to the interface of $A$, but receives the data on a narrow bus at $W
\cdot S_\text{int} = (1 \cdot 4)\;\si{\byte\per\cycle}$.  In general, this
limits the bandwidth consumption to $f \cdot W S_\text{int}$ at frequency $f$,
which is likely to be less than what the external memory can provide. To better
exploit available bandwidth, we can either read wider vectors (increase $W$) or
clock the circuit at a higher rate (increase $f$).  The former consumes more
resources, as additional logic is required to widen and narrow the data path,
but the latter is more likely to be constrained by timing constraints on the
device. 

\vspace{-0.75em} 
\subsection{Memory Striping}
\vspace{-0.25em} 
\label{sec:memory_striping}
\label{sec:striping}

When multiple memory banks with dedicated channels (e.g., multiple DRAM modules
or HBM lanes) are available, the bandwidth at which a single array is
accessed can be increased by a factor corresponding the the number of available
interfaces by striping it across memory banks. This optimization is employed by
most CPUs transparently by striping across multi-channel memory, and is commonly
known from RAID~0 configuration of disks. 

\begin{figure}[h]
  \vspace{-0.5em} 
  \begin{subfigure}{.49\columnwidth}
    \centering
    \includegraphics[width=\textwidth]{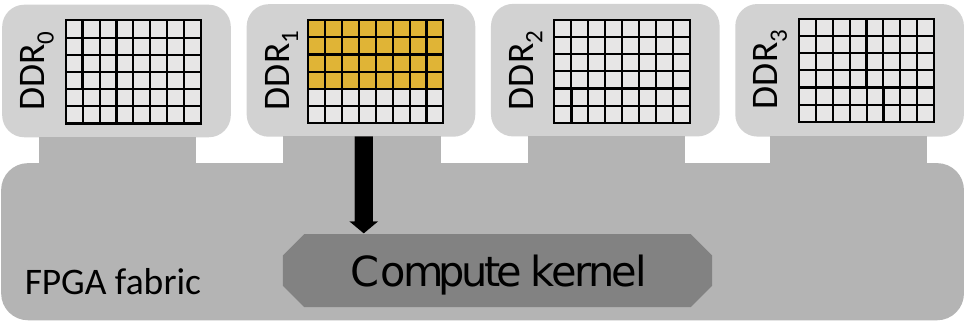}
     \vspace{-1em}
    \caption{Memory stored in a single bank.}
    \label{fig:striping_before}
  \end{subfigure}\hfill%
  \begin{subfigure}{.49\columnwidth}
    \centering
    \includegraphics[width=\textwidth]{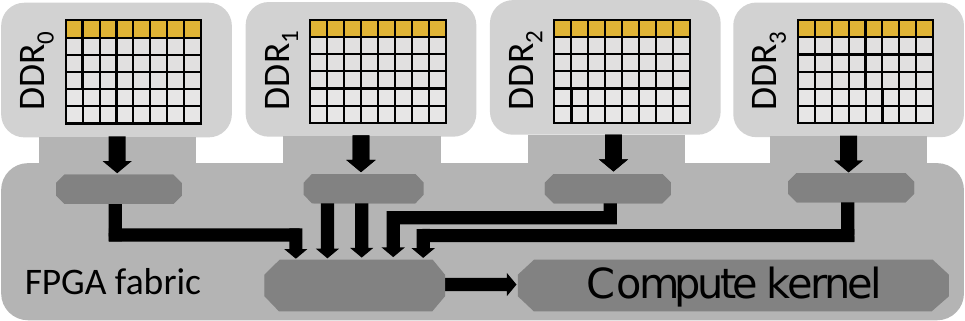}
     \vspace{-1em}
    \caption{Memory striped across four banks.}
    \label{fig:striping_after}
  \end{subfigure}\hfill%
  \vspace{-0.5em}
  \caption{Striping memory across memory banks increases available bandwidth.}
  \label{fig:striping}
  \vspace{-0.5em} 
\end{figure}

We can perform striping explicitly in HLS by inserting modules that join or
split data streams from two or more memory interfaces. Reading can be
implemented with two or more memory modules requesting memory from their
respective interfaces, pushing to FIFO buffers that are read in parallel and
combined by another module (for writing: in reverse), exposing a single data
stream to the computational kernel. This is illustrated in \figureref{striping},
where the unlabeled dark boxes in \figureref{striping_after} represent PEs
reading and combining data from the four DRAM modules.  The Intel OpenCL
compiler~\cite{altera_opencl} applies this transformation by default. 

\vspace{-0.5em} 
\subsection{Type Demotion}
\label{sec:type_demotion}

We can reduce resource and energy consumption, bandwidth requirements, and
operation latency by demoting data types to less expensive alternatives that
still meet precision requirements. This can lead to significant improvements on
architectures that are specialized for certain types, and perform poorly on
others. On traditional FPGAs there is limited native support for floating point
units. Since integer/fixed point and floating point computations on these
architectures compete for the same reconfigurable logic, using a data type with
lower resource requirements increases the total number of arithmetic operations
that can potentially be instantiated on the device. The largest benefits of type
demotion are seen in the following scenarios:
\begin{itemize}[leftmargin=*]
  \item Compute bound architectures where the data type can be changed to a type
  that occupies less of \emph{the same resources} (e.g., from $\SI{64}{\bit}$
  integers to $\SI{48}{\bit}$ integers).
  \item Compute bound architectures where the data type can be moved to a type
  that is \emph{natively} supported by the target architecture, such as single
  precision floating point on Intel's Arria~10 and Stratix~10
  devices~\cite{stratix10_native_fp}. 
  \item Bandwidth bound architectures, where performance can be improved by up
  to the same factor that the size of the data type can be reduced by. 
  \item Latency bound architectures where the data type can be reduced to a
  lower latency operation, e.g., from floating point to integer.
\end{itemize}
In the most extreme case, it has been shown that collapsing the data type of
weights and activations in deep neural networks to binary~\cite{binary_dnn} can
provide sufficient speedup for inference that the increased number of weights
makes up for the loss of precision per weight. 

\vspace{-1em} 
\section{Software Transformations in HLS}
\vspace{-0.25em} 
\label{sec:compiler_transformations}
\label{sec:software_transformations}

\begin{table}
\ssmall\sf
\setlength{\tabcolsep}{2pt}
\renewcommand{\arraystretch}{1.1}
\begin{tabular}{ p{0.02\columnwidth} p{0.02\columnwidth} p{0.91\columnwidth} }
\toprule
& \multicolumn{2}{l}{\textbf{CPU-Oriented Transformations} and how they apply to HLS codes.} \\\midrule
\multirow{22}{*}{\rotatebox[origin=c]{90}{\textbf{Directly related to HLS
transformations}}}
& \faHandORight{}\  &
  {\textbf{Loop interchange}~\cite{loop_interchange, dependence_graphs}
  is used to resolve loop-carried dependencies
  [\seclink{pipeline_enabling_transformations}].} \\
& \faHandORight{}\  &
  {\textbf{Strip-mining}~\cite{strip_mining},
  \textbf{loop tiling}~\cite{loop_tiling, dependence_graphs}, and
  \textbf{cycle shrinking}~\cite{cycle_shrinking} are central components of many
  HLS transformations [\seclink{interleaving}, \seclink{vectorization},
  \seclink{replication}, \seclink{partial_interleaving}].} \\
%
& \faHandORight{}\  &
  {\textbf{Loop distribution} and
  \textbf{loop fission}~\cite{loop_distribution, dependence_graphs}
  are used to separate differently scheduled computations to allow
  pipelining~[\seclink{dataflow}].} \\
& \faHandORight{}\  &
  \textbf{Loop fusion}~\cite{alpha, dependence_graphs, supercompilers}
  is used for merging pipelines~[\seclink{pipeline_fusion}]. \\
& \faHandORight{}\  &
  {\textbf{Loop unrolling}~\cite{loop_unrolling}
  is used to generate parallel hardware~[\seclink{vectorization},
  \seclink{replication}].} \\
& \faHandORight{}\  &
  {\textbf{Software pipelining}~\cite{software_pipelining}
  is used by HLS tools to schedule code sections according to operation
  interdependencies to form \emph{hardware} pipelines.} \\ 
& \faHandORight{}\  &
  {\textbf{Loop
  coalescing}/\textbf{flattening}/\textbf{collapsing}~\cite{loop_coalescing}
  saves pipeline drains in nested loops~[\seclink{loop_flattening}].} \\
& \faHandORight{}\  &
  {\textbf{Reduction recognition} prevents loop-carried dependencies when
  accumulating~[\seclink{interleaving}].} \\ 
& \faHandORight{}\  &
  {\textbf{Loop idiom recognition} is relevant for HLS backends, for example to
  recognize shift registers~[\seclink{fifo_buffering}] in the Intel~OpenCL
  compiler~\cite{altera_opencl}.} \\
& \faHandORight{}\  &
  {\textbf{Procedure inlining} is used to remove function call
  boundaries~[\seclink{inlining}].} \\ 
& \faHandORight{}\  &
  {\textbf{Procedure cloning} is frequently used by HLS tools when
  inlining~[\secsymbol{\ref{sec:inlining}}] to specialize each function ``call''
  with values that are known at compile-time.} \\  
& \faHandORight{}\  &
  {\textbf{Loop unswitching}~\cite{catalogue} is rarely advantageous; its
  \emph{opposite} is beneficial~[\seclink{loop_flattening},
  \seclink{pipeline_fusion}].} \\
& \faHandORight{}\  &
  {\textbf{Loop peeling} is rarely advantageous; its \textbf{opposite} is
  beneficial to allow coalescing~[\seclink{loop_flattening}].} \\
& \faHandORight{}\  &
  \textbf{SIMD transformations} is done in HLS via
  horizontal unrolling~[\seclink{vectorization}].  \\ 
& \faHandORight{}\  &
  \textbf{Short-circuiting}: while the logic for both boolean operands must
  always be instantiated in hardware, dynamically scheduling
  branches~\cite{dynamic_scheduling} can effectively ``short-circuit'' otherwise
  deep, static pipelines. \\
\midrule
\multirow{21}{*}{\rotatebox[origin=c]{90}{\textbf{Same or similar in HLS}}}
& \faHandORight{}\  &
  {\textbf{Loop-based strength reduction} \cite{strength_reduction_cocke,
  strength_reduction_bernstein, strength_reduction_steele},
  \textbf{Induction variable elimination} \cite{dragon_book},
  \textbf{Unreachable code elimination} \cite{dragon_book},
  \textbf{Useless-code elimination} \cite{dragon_book},
  \textbf{Dead-variable elimination} \cite{dragon_book},
  \textbf{Common-subexpression elimination} \cite{dragon_book},
  \textbf{Constant propagation} \cite{dragon_book},
  \textbf{Constant folding} \cite{dragon_book},
  \textbf{Copy propagation} \cite{dragon_book},
  \textbf{Forwarding substitution} \cite{dragon_book},
  \textbf{Reassociation},
  \textbf{Algebraic simplification},
  \textbf{Strength reduction},
  \textbf{Bounds reduction},
  \textbf{Redundant guard elimination}
  are all transformations that eliminate code, which is a useful step for HLS
  codes to avoid generating unnecessary hardware.} \\
& \faHandORight{}\  & \textbf{Loop-invariant code motion (hoisting)}
  \cite{dragon_book} does not save hardware in itself, but can save memory
  operations. \\
& \faHandORight{}\  &
  \textbf{Loop normalization} can be useful as an intermediate transformation.
  \\
& \faHandORight{}\  &
  \textbf{Loop reversal} \cite{dragon_book}, \textbf{array padding and
  contraction}, \textbf{scalar expansion}, and \textbf{scalar replacement} yield
  the same benefits as in software. \\
& \faHandORight{}\  &
  \textbf{Loop skewing} \cite{dragon_book} can be used in multi-dimensional
  wavefront codes. \\
& \faHandORight{}\  & \textbf{Function memoization} can be applied to HLS, using
explicit fast memory. \\
& \faHandORight{}\  &
  \textbf{Tail recursion elimination} may be useful if eliminating dynamic
  recursion can enable a code to be implemented in hardware.  \\
& \faHandORight{}\  &
  \textbf{Regular array decomposition} applies to partitioning of both on-chip/off-chip memory. \\ 
& \faHandORight{}\  &
  We do not consider transformations that apply only in a distributed setting
  (\textbf{message vectorization}, \textbf{message coalescing}, \textbf{message
  aggregation}, \textbf{collective communication}, \textbf{message pipelining},
  \textbf{guard introduction}, \textbf{redundant communication}), but they
  should be implemented in dedicated message passing hardware when
  relevant~\cite{smi}. \\
\midrule
\multirow{14}{*}{\rotatebox[origin=c]{90}{\textbf{Do not apply to HLS}}}
& \faHandORight{}\  & No use case found for \textbf{loop spreading} and
\textbf{parameter promotion}. \\
& \faHandORight{}\  &
  \textbf{Array statement scalarization}:
  No built-in vector notation in C/C++/OpenCL. \\ 
& \faHandORight{}\  &
  \textbf{Code colocation},
  \textbf{displacement minimization},
  \textbf{leaf procedure optimization}, and
  \textbf{cross-call register allocation},
  are not relevant for HLS, as there are no runtime function calls. \\ 
& \faHandORight{}\  &
  \textbf{I/O format compilation}:
  No I/O supported directly in HLS. \\ 
& \faHandORight{}\  &
  \textbf{Supercompiling}:
  is infeasible for HLS due to long synthesis times. \\
& \faHandORight{}\  &
  \textbf{Loop pushing/embedding}:
  Inlining completely is favored to allow pipelining. \\
& \faHandORight{}\  &
  {\textbf{Automatic decomposition and alignment},
  \textbf{scalar privatization}, 
  \textbf{array privatization},
  \textbf{cache alignment}, and
  \textbf{false sharing}
  are not relevant for HLS, as there is no (implicit) cache coherency protocol
  in hardware.} \\
& \faHandORight{}\  &
    \textbf{Procedure call parallelization} and \textbf{split} do not apply, as
    there are no forks in hardware. \\
& \faHandORight{}\  &
  {\textbf{Graph partitioning} only applies to explicit dataflow languages.} \\ 
& \faHandORight{}\  &
    There are no instruction sets in hardware, so \textbf{VLIW transformations}
    do not apply. \\
\bottomrule
\end{tabular}
\vspace{-1em}
  \caption{The relation of traditional CPU-oriented transformations to HLS codes.}
  \label{tab:cpu_transformations}
  \vspace{-1em}
\end{table}

In addition to the transformations described in the sections above, we include
an overview of how well-known CPU-oriented transformations apply to HLS\@, based
on the compiler transformations compiled by
Bacon~et~al.~\cite{compiler_transformations}. These transformations are included
in \tableref{cpu_transformations}, and are partitioned into three categories:
\begin{itemize}[leftmargin=*]
  \item Transformations directly relevant to the HLS transformations already
  presented here.
  \item Transformations that are the same or similar to their software
  counterparts.
  \item Transformations with little or no relevance to HLS. 
\end{itemize}
It is interesting to note that the majority of well-known transformations from
software apply to HLS\@. This implies that we can leverage much of decades of
research into high-performance computing transformations to also optimize
hardware programs, including many that can be applied \emph{directly} (i.e.,
without further adaptation to HLS) to the imperative source code or intermediate
representation before synthesizing for hardware.
We stress the importance of support for these pre-hardware generation
transformations in HLS compilers, as they lay the foundation for the
hardware-specific transformations proposed here.

\vspace{-1em} 
\section{End-to-End Examples}
\vspace{-0.25em} 
\label{sec:evaluation}
\label{sec:end_to_end_examples}
\label{sec:application_examples}

To showcase the transformations presented in this work and provide a
``hands-on'' opportunity for seeing HLS optimizations applied in practice, we
will describe the optimization process on a sample set of classical HPC kernels,
available as open source repositories on
github\footnote{\vspace{-0.5em}\url{https://github.com/spcl?q=hls}}.
These kernels are written in C++ for Xilinx Vivado~HLS~\cite{autopilot} with
hlslib~\cite{hlslib} extensions, and are built and run using the Xilinx Vitis
environment.
For each example, we will describe the sequence of transformations applied, and
give the resulting performance at each major stage.

The included benchmarks were run on an Alveo U250 board, which houses a Xilinx
UltraScale+ XCU250-FIGD2104-2L-E FPGA and four
$\SI[per-mode=symbol]{2400}{\mega\transfer\per\second}$ DDR4 banks (we utilize
1-2 banks for the examples here). The chip consists of four almost identical
chiplets with limited interconnect between them, where each chiplet is connected
to one of the DDR4 pinouts. This multi-chiplet design allows more resources
($\SI{1728}{}\text{K}$ LUTs and $\SI{12288}{}$ DSPs), but poses challenges for
the routing process, which impedes the achievable clock rate and resource
utilization for a monolithic kernel attempting to span the full chip.
Kernels were compiled for the xilinx\_u250\_xdma\_201830\_2 shell with
Vitis~2019.2 and executed with version 2.3.1301 of the Xilinx Runtime (XRT).
All benchmarks are included in \figureref{transformation_speedup}, and the
resource utilization of each kernel is shown in
\figureref{transformation_resources}. 

\vspace{-0.5em} 
\subsection{Stencil Code}
\vspace{-0.25em} 

\begin{figure}
  \begin{minipage}{\columnwidth}
    \centering
    \includegraphics[width=.99\columnwidth]{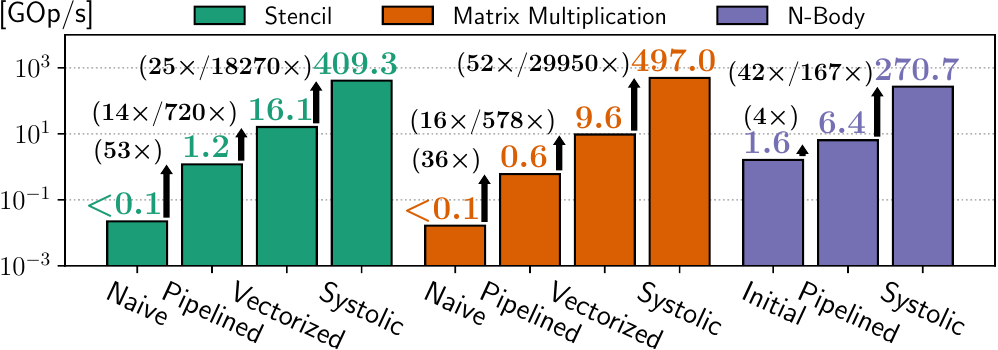}
    \vspace{-0.75em}
    \caption{Performance progression of kernels when applying transformations.
    Parentheses show speedup over previous version, and cumulative speedup.}
    \label{fig:transformation_speedup}
  \end{minipage}

  \vspace{0.5em}

  \begin{minipage}{\columnwidth}
    \centering
    \includegraphics[width=.99\columnwidth]{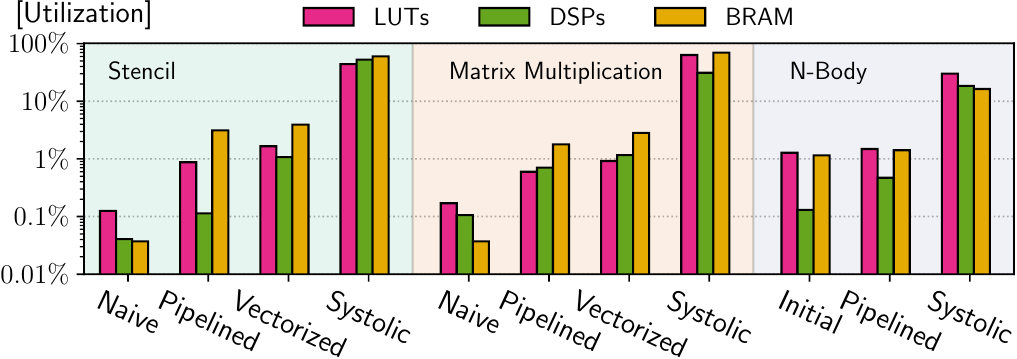}
    \vspace{-0.75em}
    \caption{Resource usage of kernels from \figureref{transformation_speedup}
    as fractions of available resources. The maxima are taken as
    $\SI{1728}{}\text{K}$ LUTs, $\SI{12288}{}$ DSPs, and $\SI{2688}{}$ BRAM.}
    \vspace{-0.5em} 
    \label{fig:transformation_resources}
  \end{minipage}
\end{figure}

Stencil codes are a popular target for FPGA acceleration in HPC, due to their
regular access pattern, intuitive buffering scheme, and potential for creating
large systolic array designs~\cite{opencl_stencil_optimization_zohouri}.
We show the optimization of a 4-point 2D stencil based on \listingref{stencil}.
Benchmarks are shown in \figureref{transformation_speedup}, and use single
precision floating point, iterating over a $\SI{8192}{}{\times}\SI{8192}{}$
domain.  We first measure a naive implementation, where all neighboring cells
are accessed directly from the input array, which results in no data reuse and
heavy interface contention on the input array. We then apply the following
optimization steps:  
\begin{enumerate}[leftmargin=*]
  \item Delay buffers~[\seclink{fifo_buffering}] are added to store two rows of
  the domain (see \listingref{stencil_streams}), removing interface contention
  on the memory bus and achieving perfect spatial data reuse.
  \item Spatial locality is exploited by introducing
  vectorization~[\seclink{vectorization}]. To efficiently use memory bandwidth,
  we use memory extraction~[\seclink{memory_access_extraction}],
  buffering~[\seclink{memory_buffering}], and
  striping~[\seclink{memory_striping}] from two DDR banks. 
  \item To exploit temporal locality, we replicate the vectorized PE by vertical
  unrolling~[\seclink{replication}] and stream~[\seclink{dataflow}] between
  them (\listingref{stencil_dataflow}). The domain is tiled~[\seclink{tiling}]
  to limit fast memory usage.
\end{enumerate}

\noindent Enabling pipelining with delay buffers allows the kernel to throughput
${\sim}1$ cell per cycle. Improving the memory performance to add vectorization
(using $W=\SI{16}{\operands\per\cycle}$ for the kernel) exploits spatial
locality through additional bandwidth usage. The vertical unrolling and dataflow
step scales the design to exploit available hardware resources on the chip,
until limited by placement and routing. The final implementation is available on
github\footnote{\vspace{-0.5em}\url{https://github.com/spcl/stencil_hls/}}.

\vspace{-0.5em} 
\subsection{Matrix Multiplication Code}
\vspace{-0.25em} 

We consider the optimization process of a matrix multiplication kernel using
transformations presented here. Benchmark for $\SI{8192}{}{\times}\SI{8192}{}$
matrices across stages of optimization are shown in
\figureref{transformation_speedup}. Starting from a naive implementation
(\listingref{gemm_naive}), the following optimization stages were applied:
\begin{enumerate}[leftmargin=*]
  \item We transpose the iteration space~[\seclink{transposition}], removing the
  loop-carried dependency on the accumulation register, and extract the memory
  accesses~[\seclink{memory_access_extraction}], vastly improving spatial
  locality. The buffering, streaming and writing
  phases are fused~[\seclink{pipeline_fusion}], allowing us to coalesce the
  three nested loops~[\seclink{loop_flattening}].
  \item In order to increase spatial parallelism, we vectorize accesses to
  $\matr{B}$ and $\matr{C}$~[\seclink{vectorization}].
  \item To scale up the design, we vertically unroll by buffering multiple
  values of $\matr{A}$, applying them to streamed in values of $\matr{B}$ in
  parallel~[\seclink{replication}]. To avoid high fan-out, we partition buffered
  elements of $\matr{A}$ into processing elements~[\seclink{dataflow}] arranged
  in a systolic array architecture.  Finally, the horizontal domain is tiled to
  accommodate arbitrarily large matrices with finite buffer space. 
\end{enumerate}

\noindent Allowing pipelining and regularizing the memory access pattern results
in a throughput of ${\sim}1$ cell per cycle. Vectorization multiplies the
performance by $W$, set to $16$ in the benchmarked kernel. The performance of
the vertically unrolled dataflow kernel is only limited by placement and routing
due to high resource usage on the chip.  The final implementation achieves
state-of-the-art performance on the target
architecture~\cite{matrix_multiplication_definelicht}, and is available on
github\footnote{\vspace{-0.5em}\url{https://github.com/spcl/gemm_hls}}. 

\vspace{-0.5em} 
\subsection{N-Body Code}
\vspace{-0.25em} 

Finally, we show an N-body code in 3 dimensions, using single precision floating
point types and iterating over $\SI{16128}{}$ bodies. Since Vivado~HLS does not
allow memory accesses of a width that is not a power of two,  memory
extraction~[\seclink{memory_access_extraction}] and
buffering~[\seclink{memory_buffering}] was included in the first stage, to
support 3-vectors of velocity. We then performed the following transformations:
\begin{enumerate}[leftmargin=*]
  \item The loop-carried dependency on the acceleration accumulation is resolved
  by applying tiled accumulation
  interleaving~[\seclink{nested_accumulation_interleaving}], pipelining across
  $T{\geq}L_+$ different resident particles applied to particles streamed in.
  \item To scale up the performance, we further multiply the number of resident
  particles, this time replicating compute through vertical
  unrolling~[\seclink{replication}] of the outer loop into $P$ parallel
  processing element arranged in a systolic array. Each element holds $T$
  resident particles, and particles are streamed~[\seclink{dataflow}] through
  the PEs. 
\end{enumerate}
The second stage gains a factor of $4\times$ corresponding to the latency of the
interleaved accumulation, followed by a factor of $42\times$ from unrolling
units across the chip.
$T{\geq}L_+$ can be used to regulate the arithmetic intensity of the kernel. The
bandwidth requirements can be reduced further by storing more resident particles
on the chip, scaling up to the full fast memory usage of the FPGA\@. The tiled
accumulation interleaving transformation thus enables not just pipelining of the
compute, but also minimization of I/O. 
The optimized implementation is available on
github\footnote{\vspace{-0.5em}\url{https://github.com/spcl/nbody_hls}}.

\vspace{0.25em}

These examples demonstrate the impact of different transformations on a
reconfigurable hardware platform. In particular, enabling pipelining,
regularizing memory accesses, and vertical unrolling are shown to be central
components of scalable hardware architectures. The dramatic speedups over naive
codes also emphasize that HLS tools do not yield competitive performance out of
the box, making it critical to perform further transformations. For additional
examples of optimizing HLS codes, we refer to the numerous works applying HLS
optimizations listed below.

\vspace{-0.75em} 
\section{Related Work}
\vspace{-0.25em} 
\label{sec:related_work}

\hspace{\parindent} \textbf{\emph{Optimized applications.}}
Much work has been done in optimizing C/C++/OpenCL HLS codes for FPGA, such as
stencils~\cite{opencl_stencil_optimization_zohouri,
stencil_opencl_optimization_waidyasooriya, opencl_stencil_optimization_jia,
opencl_stencil_optimizations_weller, opendwarfs}, deep neural
networks~\cite{opencl_dnn_0, opencl_dnn_1, hls_dnn_0, finn_r, binary_dnn},
matrix multiplication~\cite{matrix_multiplication_dhollander, opendwarfs,
matrix_multiplication_definelicht, cannon}, graph
processing~\cite{fpga_graph_survey, substream_centric},
networking~\cite{vivado_dataflow}, light propagation for cancer
treatment~\cite{light_propagation}, and protein
sequencing~\cite{opencl_smith_waterman_0, opencl_smith_waterman_1}. These works
optimize the respective applications using transformations described here, such
as delay buffering, random access buffering, vectorization, vertical unrolling,
tiling for on-chip memory, and dataflow.  

\textbf{\emph{Transformations.}}
Zohouri~et~al.~\cite{opencl_optimization} use the Rodinia benchmark to evaluate
the performance of OpenCL codes targeting FPGAs, employing optimizations such as
SIMD vectorization, sliding-window buffering, accumulation interleaving, and
compute unit replication across multiple kernels. We present a generalized
description of a superset of these transformations, along with concrete code
examples that show how they are applied in practice.
The DaCe framework~\cite{dace} exploits
information on explicit dataflow and control flow to perform a wide range of
transformations, and code generates efficient HLS code using vendor-specific
pragmas and primitives.
Kastner~et~al.~\cite{parallel_hls_book} go through the implementation of many
HLS codes in Vivado~HLS, focusing on algorithmic optimizations.
da~Silva~et~al.~\cite{module_per_object} explore using modern C++ features to
capture HLS concepts in a high-level fashion.
Lloyd~et~al.~\cite{opencl_host_optimizations} describe optimizations specific to
Intel~OpenCL, and include a variant of memory access extraction, as well as the
single-loop accumulation variant of accumulation interleaving.

\textbf{\emph{Directive-based frameworks.}}
High-level, directive-based frameworks such as OpenMP and OpenACC have been
proposed as alternative abstractions for generating FPGA kernels.
Leow~et~al.~\cite{openmp_for_fpga_0} implement an FPGA code generator from
OpenMP pragmas, primarily focusing on correctness in implementing a range of
OpenMP pragmas.
Lee~et~al.~\cite{openacc_to_fpga} present an OpenACC to OpenCL compiler, using
Intel~OpenCL as a backend. The authors implement horizontal and vertical
unrolling, pipelining and dataflow by introducing new OpenACC clauses.
Papakonstantinou~et~al.~\cite{cuda_to_fpga} generate HLS code for FPGA from
directive-annotated CUDA code.

\textbf{\emph{Optimizing HLS compilers.}}
Mainstream HLS compilers automatically apply many of the well-known software
transformations in \tableref{cpu_transformations}~\cite{liquid_metal,
spark_optimizing_hls_0, spark_optimizing_hls_1}, but can also employ more
advanced FPGA transformations.
Intel~OpenCL~\cite{altera_opencl} performs memory access extraction into ``load
store units'' (LSUs), does memory striping between DRAM banks, and detects and
auto-resolves some buffering and accumulation patterns.
The proprietary Merlin Compiler~\cite{merlin_compiler} uses high-level
acceleration directives to automatically perform some of the transformations
described here, as source-to-source transformations to underlying HLS code.
Polyhedral compilation is a popular framework for optimizing CPU and GPU loop
nests~\cite{polly}, and has also been applied to HLS for FPGA for optimizing
data reuse~\cite{polyhedral_hls_1}. Such techniques may prove valuable in
automating, e.g., memory extraction and tiling transformations. 
While most HLS compilers rely strictly on static scheduling,
Dynamatic~\cite{dynamic_scheduling} considers dynamically scheduling state
machines and pipelines to allow reducing the number of stages executed at
runtime. 

\textbf{\emph{Domain-specific frameworks.}}
Implementing programs in domain specific languages (DSLs) can make it easier to
detect and exploit opportunities for advanced transformations.
Darkroom~\cite{darkroom} generates optimized HDL for image processing codes, and
the popular image processing framework Halide~\cite{halide} has been extended to
support FPGAs~\cite{halide_fpga, hetereohalide}.
Luzhou~et~al.~\cite{stencil_dsl} and StencilFlow~\cite{stencilflow} propose
frameworks for generating stencil codes for FPGAs. These frameworks rely on
optimizations such as delay buffering, dataflow, and vertical unrolling, which
we cover here.
Using DSLs to compile to structured HLS code can be a viable approach to
automating a wide range of transformations, as proposed by
Koeplinger~et~al.~\cite{parallel_patterns}, and the FROST~\cite{frost} DSL
framework. 


\textbf{\emph{Other approaches.}}
There are other approaches than C/C++/OpenCL-based HLS languages to addressing
the productivity issues of hardware design:
Chisel/FIRRTL~\cite{chisel, firrtl} maintains the paradigm of behavioral
programming known from RTL, but provides modern language and compiler features.
This caters to developers who are already familiar with hardware design, but
wish to use a more expressive language.
In the Maxeler ecosystem~\cite{maxcompiler}, kernels are described using a
Java-based language, but rather than transforming imperative code into a
behavioral equivalent, the language provides a DSL of hardware concepts that are
instantiated using object-oriented interfaces.
By constraining the input, this encourages developers to write code that maps
well to hardware, but requires learning a new language exclusive to the Maxeler
ecosystem.

\vspace{-1.25em}
\section{Toolflow of Xilinx vs. Intel}
\vspace{-0.25em}

When choosing a toolflow to start designing hardware with HLS, it is useful to
understand two distinct approaches by the two major vendors: Intel OpenCL wishes
to enable \emph{writing accelerators using software}, making an effort to
abstract away low-level details about the hardware, and present a high-level
view to the programmer; whereas Xilinx' Vivado~HLS provides \emph{a more
productive way of writing hardware}, by means of a familiar software language.
Xilinx uses OpenCL as a vehicle to interface between FPGA and host, but
implements the OpenCL compiler itself as a thin wrapper around the C++ compiler,
whereas Intel embraces the OpenCL paradigm as their frontend (although they
encourage writing single work item kernels~\cite{intel_opencl_best_practices},
effectively preventing reuse of OpenCL kernels written for GPU).

Vivado HLS has a stronger coupling between the HLS source code and the generated
hardware. This requires the programmer to write more annotations and boilerplate
code, but can also give them stronger feeling of control.
Conversely, the Intel OpenCL compiler presents convenient abstracted views,
saves boilerplate code (e.g., by automatically pipelining sections), and
implements efficient substitutions by detecting common patterns in the source
code (e.g., to automatically perform memory
extraction~[\seclink{memory_access_extraction}]).  The downside is that
developers end up struggling to write or generate code in a way that is
recognized by the tool's ``black magic'', in order to achieve the desired
result. 
Finally, Xilinx' choice to allow C++ gives Vivado HLS an edge in expressibility,
as (non-virtual) objects and templating turns out to be a useful tool for
abstracting and extending the language~\cite{hlslib}. Intel offers a C++-based
HLS compiler, but does not (as of writing) support direct interoperability with
the OpenCL-driven accelerator flow. 

\vspace{-0.75em} 
\section{Conclusion}
\vspace{-0.25em} 
\label{sec:conclusion}

The transformations known from software are insufficient to optimize HPC kernels
targeting spatial computing systems.
We have proposed a new set of optimizing transformations that enable efficient
and scalable hardware architectures, and can be applied directly to the source
code by a performance engineer, or automatically by an optimizing compiler.
Performance and compiler engineers can benefit from these guidelines,
transformations, and the presented cheat~sheet as a common toolbox for
developing high performance hardware using HLS.

\ifoptionfinal{%
\vspace{-0.75em} 
\section*{Acknowledgements}
\vspace{-0.25em} 
\label{sec:acknowledgements}
This work was supported by the European Research Council under the European
Union's Horizon 2020 programme (grant agreement DAPP, No. 678880). The authors
wish to thank Xilinx and Intel for helpful discussions; Xilinx for generous
donations of software, hardware, and access to the Xilinx Adaptive Compute
Cluster (XACC) at ETH Zurich; the Swiss National Supercomputing Center (CSCS)
for providing computing infrastructure; and Tal Ben-Nun for valuable feedback
on iterations of this manuscript.
}{%
\vspace{-2em} 
}


\bibliographystyle{IEEEtran}
\bibliography{HLS_Transformations}

\begin{thebibliography}{100}
\providecommand{\url}[1]{#1}
\csname url@samestyle\endcsname
\providecommand{\newblock}{\relax}
\providecommand{\bibinfo}[2]{#2}
\providecommand{\BIBentrySTDinterwordspacing}{\spaceskip=0pt\relax}
\providecommand{\BIBentryALTinterwordstretchfactor}{4}
\providecommand{\BIBentryALTinterwordspacing}{\spaceskip=\fontdimen2\font plus
\BIBentryALTinterwordstretchfactor\fontdimen3\font minus
  \fontdimen4\font\relax}
\providecommand{\BIBforeignlanguage}[2]{{%
\expandafter\ifx\csname l@#1\endcsname\relax
\typeout{** WARNING: IEEEtran.bst: No hyphenation pattern has been}%
\typeout{** loaded for the language `#1'. Using the pattern for}%
\typeout{** the default language instead.}%
\else
\language=\csname l@#1\endcsname
\fi
#2}}
\providecommand{\BIBdecl}{\relax}
\BIBdecl

\bibitem{hitting_the_power_wall}
W.~A. Wulf and S.~A. McKee, ``Hitting the memory wall: implications of the
  obvious,'' \emph{SIGARCH}, 1995.

\bibitem{computings_energy_problem}
M.~Horowitz, ``Computing's energy problem (and what we can do about it),'' in
  \emph{ISSCC}, 2014.

\bibitem{second_opinion_on_dataflow}
D.~D. Gajski \emph{et~al.}, ``A second opinion on data flow machines and
  languages,'' \emph{Computer}, 1982.

\bibitem{wheres_the_beef}
S.~Sirowy and A.~Forin, ``Where's the beef? why {FPGAs} are so fast,'' \emph{MS
  Research}, 2008.

\bibitem{platform_comparison}
A.~R. Brodtkorb \emph{et~al.}, ``State-of-the-art in heterogeneous computing,''
  \emph{Scientific Programming}, 2010.

\bibitem{rng_comparison}
D.~B. Thomas \emph{et~al.}, ``A comparison of {CPUs}, {GPUs}, {FPGAs}, and
  massively parallel processor arrays for random number generation,'' in
  \emph{FPGA}, 2009.

\bibitem{fpga_for_the_masses}
D.~Bacon \emph{et~al.}, ``{FPGA} programming for the masses,'' \emph{CACM},
  2013.

\bibitem{hls_past_present_future}
G.~Martin and G.~Smith, ``High-level synthesis: Past, present, and future,''
  \emph{D\&T}, 2009.

\bibitem{hls_for_fpgas}
J.~Cong \emph{et~al.}, ``High-level synthesis for {FPGAs}: From prototyping to
  deployment,'' \emph{TCAD}, 2011.

\bibitem{hls_survey_nane}
R.~Nane \emph{et~al.}, ``A survey and evaluation of {FPGA} high-level synthesis
  tools,'' \emph{TCAD}, 2016.

\bibitem{hls_survey_meeus}
W.~Meeus \emph{et~al.}, ``An overview of today's high-level synthesis tools,''
  \emph{DAEM}, 2012.

\bibitem{autopilot}
Z.~Zhang \emph{et~al.}, ``{AutoPilot}: A platform-based {ESL} synthesis
  system,'' in \emph{High-Level Synthesis}, 2008.

\bibitem{intel_hls}
{Intel High-Level Synthesis ({HLS}) Compiler}.
  \url{https://www.intel.com/content/www/us/en/software/programmable/quartus-prime/hls-compiler.html}.
  Accessed May 15, 2020.

\bibitem{legup}
A.~Canis \emph{et~al.}, ``{LegUp}: High-level synthesis for {FPGA}-based
  processor/accelerator systems,'' in \emph{FPGA}, 2011.

\bibitem{catapult_c}
{Mentor Graphics}. {Catapult} high-level synthesis.
  \url{https://www.mentor.com/hls-lp/catapult-high-level-synthesis/c-systemc-hls}.
  Accessed May 15, 2020.

\bibitem{bambu}
C.~Pilato \emph{et~al.}, ``{Bambu}: A modular framework for the high level
  synthesis of memory-intensive applications,'' in \emph{FPL}, 2013.

\bibitem{dwarv}
R.~Nane \emph{et~al.}, ``{DWARV} 2.0: A {CoSy}-based {C}-to-{VHDL} hardware
  compiler,'' in \emph{FPL}, 2012.

\bibitem{opencl_owaida}
M.~Owaida \emph{et~al.}, ``Synthesis of platform architectures from {OpenCL}
  programs,'' in \emph{FCCM}, 2011.

\bibitem{altera_opencl}
T.~Czajkowski \emph{et~al.}, ``From {OpenCL} to high-performance hardware on
  {FPGAs},'' in \emph{FPL}, 2012.

\bibitem{bluespec}
R.~Nikhil, ``{Bluespec} system {Verilog}: efficient, correct {RTL} from high
  level specifications,'' in \emph{MEMOCODE}, 2004.

\bibitem{lime}
J.~Auerbach \emph{et~al.}, ``{Lime}: A {Java}-compatible and synthesizable
  language for heterogeneous architectures,'' in \emph{OOPSLA}, 2010.

\bibitem{liquid_metal}
------, ``A compiler and runtime for heterogeneous computing,'' in \emph{DAC},
  2012.

\bibitem{esterel}
J.~Hammarberg and S.~Nadjm-Tehrani, ``Development of safety-critical
  reconfigurable hardware with {Esterel},'' \emph{FMICS}, 2003.

\bibitem{streamsc}
M.~B. Gokhale \emph{et~al.}, ``Stream-oriented {FPGA} computing in the
  {Streams-C} high level language,'' in \emph{FCCM}, 2000.

\bibitem{compiler_transformations}
D.~F. Bacon \emph{et~al.}, ``Compiler transformations for high-performance
  computing,'' \emph{CSUR}, 1994.

\bibitem{gpu_optimization}
S.~Ryoo \emph{et~al.}, ``Optimization principles and application performance
  evaluation of a multithreaded {GPU} using {CUDA},'' in \emph{PPoPP}, 2008.

\bibitem{stencil_pde}
G.~D. Smith, \emph{Numerical solution of partial differential equations: finite
  difference methods}, 1985.

\bibitem{stencil_electrodynamics}
A.~Taflove and S.~C. Hagness, ``Computational electrodynamics: The
  finite-difference time-domain method,'' 1995.

\bibitem{stencil_cfd}
C.~A. Fletcher, \emph{Computational Techniques for Fluid Dynamics 2}, 1988.

\bibitem{darkroom}
J.~Hegarty \emph{et~al.}, ``Darkroom: compiling high-level image processing
  code into hardware pipelines.'' \emph{TOG}, 2014.

\bibitem{halide}
J.~Ragan-Kelley \emph{et~al.}, ``{Halide}: A language and compiler for
  optimizing parallelism, locality, and recomputation in image processing
  pipelines,'' in \emph{PLDI}, 2013.

\bibitem{dnn_performance_survey}
T.~Ben-Nun and T.~Hoefler, ``Demystifying parallel and distributed deep
  learning: An in-depth concurrency analysis,'' \emph{CSUR}, 2019.

\bibitem{fpga_dnn_survey}
G.~Lacey \emph{et~al.}, ``Deep learning on {FPGAs}: Past, present, and
  future,'' \emph{arXiv:1602.04283}, 2016.

\bibitem{binary_dnn}
M.~Courbariaux \emph{et~al.}, ``Binarized neural networks: Training deep neural
  networks with weights and activations constrained to +1 or -1,''
  \emph{arXiv:1602.02830}, 2016.

\bibitem{hls_dnn_0}
Y.~Umuroglu \emph{et~al.}, ``{FINN}: A framework for fast, scalable binarized
  neural network inference,'' in \emph{FPGA}, 2017.

\bibitem{finn_r}
M.~Blott \emph{et~al.}, ``{FINN-R}: An end-to-end deep-learning framework for
  fast exploration of quantized neural networks,'' \emph{TRETS}, 2018.

\bibitem{reverse_time_migration}
H.~Fu and R.~G. Clapp, ``Eliminating the memory bottleneck: An {FPGA}-based
  solution for 3d reverse time migration,'' in \emph{FPGA}, 2011.

\bibitem{opencl_stencil_optimization_zohouri}
H.~R. Zohouri \emph{et~al.}, ``Combined spatial and temporal blocking for
  high-performance stencil computation on {FPGAs} using {OpenCL},'' in
  \emph{FPGA}, 2018.

\bibitem{stencil_opencl_optimization_waidyasooriya}
H.~M. Waidyasooriya \emph{et~al.}, ``{OpenCL}-based {FPGA}-platform for stencil
  computation and its optimization methodology,'' \emph{TPDS}, May 2017.

\bibitem{opencl_stencil_optimization_jia}
Q.~Jia and H.~Zhou, ``Tuning stencil codes in {OpenCL} for {FPGAs},'' in
  \emph{ICCD}, 2016.

\bibitem{stencil_runtime_reconfig_0}
X.~Niu \emph{et~al.}, ``Exploiting run-time reconfiguration in stencil
  computation,'' in \emph{FPL}, 2012.

\bibitem{stencil_runtime_reconfig_1}
------, ``Dynamic stencil: Effective exploitation of run-time resources in
  reconfigurable clusters,'' in \emph{FPT}, 2013.

\bibitem{sliding_window_fpga}
J.~Fowers \emph{et~al.}, ``A performance and energy comparison of {FPGAs},
  {GPUs}, and multicores for sliding-window applications,'' in \emph{FPGA},
  2012.

\bibitem{stencilflow}
J.~de~Fine~Licht \emph{et~al.}, ``{StencilFlow}: Mapping large stencil programs
  to distributed spatial computing systems,'' in \emph{CGO}, 2021.

\bibitem{bram_shuffling}
X.~Chen \emph{et~al.}, ``On-the-fly parallel data shuffling for graph
  processing on {OpenCL}-based {FPGAs},'' in \emph{FPL}, 2019.

\bibitem{light_propagation}
T.~Young-Schultz \emph{et~al.}, ``Using {OpenCL} to enable software-like
  development of an {FPGA}-accelerated biophotonic cancer treatment
  simulator,'' in \emph{FPGA}, 2020.

\bibitem{dependence_graphs}
D.~J. Kuck \emph{et~al.}, ``Dependence graphs and compiler optimizations,'' in
  \emph{POPL}, 1981.

\bibitem{hlslib}
J.~de~Fine~Licht and T.~Hoefler, ``{hlslib}: Software engineering for hardware
  design,'' \emph{arXiv:1910.04436}, 2019.

\bibitem{opencl_smith_waterman_0}
S.~O. Settle, ``High-performance dynamic programming on {FPGAs} with
  {OpenCL},'' in \emph{HPEC}, 2013.

\bibitem{matrix_multiplication_definelicht}
J.~de~Fine~Licht \emph{et~al.}, ``Flexible communication avoiding matrix
  multiplication on {FPGA} with high-level synthesis,'' in \emph{FPGA}, 2020.

\bibitem{multi_fpga_stencil}
K.~{Sano} \emph{et~al.}, ``Multi-{FPGA} accelerator for scalable stencil
  computation with constant memory bandwidth,'' \emph{TPDS}, 2014.

\bibitem{systolic_arrays}
H.~Kung and C.~E. Leiserson, ``Systolic arrays (for {VLSI}),'' in \emph{Sparse
  Matrix Proceedings}, 1978.

\bibitem{stencil_dsl}
W.~Luzhou \emph{et~al.}, ``Domain-specific language and compiler for stencil
  computation on fpga-based systolic computational-memory array,'' in
  \emph{ARC}, 2012.

\bibitem{galerkin}
T.~Kenter \emph{et~al.}, ``{OpenCL}-based {FPGA} design to accelerate the nodal
  discontinuous {Galerkin} method for unstructured meshes,'' in \emph{FCCM},
  2018.

\bibitem{polly}
T.~Grosser \emph{et~al.}, ``Polly -- performing polyhedral optimizations on a
  low-level intermediate representation,'' \emph{PPL}, 2012.

\bibitem{stratix10_native_fp}
U.~Sinha, ``Enabling impactful {DSP} designs on {FPGAs} with hardened
  floating-point implementation,'' \emph{Altera White Paper}, 2014.

\bibitem{loop_interchange}
J.~R. Allen and K.~Kennedy, ``Automatic loop interchange,'' in \emph{SIGPLAN},
  1984.

\bibitem{strip_mining}
M.~Weiss, ``Strip mining on {SIMD} architectures,'' in \emph{ICS}, 1991.

\bibitem{loop_tiling}
M.~D. Lam \emph{et~al.}, ``The cache performance and optimizations of blocked
  algorithms,'' 1991.

\bibitem{cycle_shrinking}
C.~D. Polychronopoulos, ``Advanced loop optimizations for parallel computers,''
  in \emph{ICS}, 1988.

\bibitem{loop_distribution}
D.~J. Kuck, ``A survey of parallel machine organization and programming,''
  \emph{CSUR}, Mar. 1977.

\bibitem{alpha}
A.~P. Yershov, ``{ALPHA} -- an automatic programming system of high
  efficiency,'' \emph{J. ACM}, 1966.

\bibitem{supercompilers}
M.~J. Wolfe, ``Optimizing supercompilers for supercomputers,'' Ph.D.
  dissertation, 1982.

\bibitem{loop_unrolling}
J.~J. Dongarra and A.~R. Hinds, ``Unrolling loops in {Fortran},''
  \emph{Software: Practice and Experience}, 1979.

\bibitem{software_pipelining}
M.~Lam, ``Software pipelining: An effective scheduling technique for {VLIW}
  machines,'' in \emph{PLDI}, 1988.

\bibitem{loop_coalescing}
C.~D. Polychronopoulos, ``Loop coalescing: A compiler transformation for
  parallel machines,'' Tech. Rep., 1987.

\bibitem{catalogue}
F.~E. Allen and J.~Cocke, \emph{A catalogue of optimizing transformations},
  1971.

\bibitem{dynamic_scheduling}
L.~Josipovi{\'c} \emph{et~al.}, ``Dynamically scheduled high-level synthesis,''
  in \emph{FPGA}, 2018.

\bibitem{strength_reduction_cocke}
J.~Cocke and K.~Kennedy, ``An algorithm for reduction of operator strength,''
  \emph{CACM}, 1977.

\bibitem{strength_reduction_bernstein}
R.~Bernstein, ``Multiplication by integer constants,'' \emph{Softw. Pract.
  Exper.}, 1986.

\bibitem{strength_reduction_steele}
G.~L. Steele, ``Arithmetic shifting considered harmful,'' \emph{ACM SIGPLAN
  Notices}, 1977.

\bibitem{dragon_book}
A.~V. Aho \emph{et~al.}, ``Compilers, principles, techniques,'' \emph{Addison
  Wesley}, 1986.

\bibitem{smi}
T.~De~Matteis \emph{et~al.}, ``Streaming message interface: High-performance
  distributed memory programming on reconfigurable hardware,'' in \emph{SC},
  2019.

\bibitem{opencl_stencil_optimizations_weller}
D.~Weller \emph{et~al.}, ``Energy efficient scientific computing on {FPGAs}
  using {OpenCL},'' in \emph{FPGA}, 2017.

\bibitem{opendwarfs}
A.~Verma \emph{et~al.}, ``Accelerating workloads on {FPGAs} via {OpenCL}: A
  case study with opendwarfs,'' Tech. Rep., 2016.

\bibitem{opencl_dnn_0}
N.~Suda \emph{et~al.}, ``Throughput-optimized {OpenCL}-based {FPGA} accelerator
  for large-scale convolutional neural networks,'' in \emph{FPGA}, 2016.

\bibitem{opencl_dnn_1}
J.~Zhang and J.~Li, ``Improving the performance of {OpenCL}-based {FPGA}
  accelerator for convolutional neural network,'' in \emph{FPGA}, 2017.

\bibitem{matrix_multiplication_dhollander}
E.~H. D'Hollander, ``High-level synthesis optimization for blocked
  floating-point matrix multiplication,'' \emph{SIGARCH}, 2017.

\bibitem{cannon}
P.~Gorlani \emph{et~al.}, ``{OpenCL} implementation of {Cannon's} matrix
  multiplication algorithm on {Intel} {Stratix 10} {FPGAs},'' in \emph{ICFPT},
  2019.

\bibitem{fpga_graph_survey}
M.~Besta \emph{et~al.}, ``Graph processing on {FPGAs}: Taxonomy, survey,
  challenges,'' \emph{arXiv:1903.06697}, 2019.

\bibitem{substream_centric}
------, ``Substream-centric maximum matchings on {FPGA},'' in \emph{FPGA},
  2019.

\bibitem{vivado_dataflow}
H.~Eran \emph{et~al.}, ``Design patterns for code reuse in {HLS} packet
  processing pipelines,'' in \emph{FCCM}, 2019.

\bibitem{opencl_smith_waterman_1}
E.~Rucci \emph{et~al.}, ``{Smith-Waterman} protein search with {OpenCL} on an
  {FPGA},'' in \emph{Trustcom/BigDataSE/ISPA}, 2015.

\bibitem{opencl_optimization}
H.~R. Zohouri \emph{et~al.}, ``Evaluating and optimizing {OpenCL} kernels for
  high performance computing with {FPGAs},'' in \emph{SC}, 2016.

\bibitem{dace}
T.~Ben-Nun \emph{et~al.}, ``Stateful dataflow multigraphs: A data-centric model
  for performance portability on heterogeneous architectures,'' in \emph{SC},
  2019.

\bibitem{parallel_hls_book}
R.~Kastner \emph{et~al.}, ``Parallel programming for {FPGAs},''
  \emph{arXiv:1805.03648}, 2018.

\bibitem{module_per_object}
J.~S. da~Silva \emph{et~al.}, ``{Module-per-Object}: a human-driven methodology
  for {C++}-based high-level synthesis design,'' in \emph{FCCM}, 2019.

\bibitem{opencl_host_optimizations}
T.~Lloyd \emph{et~al.}, ``A case for better integration of host and target
  compilation when using {OpenCL} for {FPGAs},'' in \emph{FSP}, 2017.

\bibitem{openmp_for_fpga_0}
Y.~Y. Leow \emph{et~al.}, ``Generating hardware from {OpenMP} programs,'' in
  \emph{FPT}, 2006.

\bibitem{openacc_to_fpga}
S.~Lee \emph{et~al.}, ``{OpenACC} to {FPGA}: A framework for directive-based
  high-performance reconfigurable computing,'' in \emph{IPDPS}, 2016.

\bibitem{cuda_to_fpga}
A.~Papakonstantinou \emph{et~al.}, ``{FCUDA}: Enabling efficient compilation of
  {CUDA} kernels onto {FPGAs},'' in \emph{SASP}, 2009.

\bibitem{spark_optimizing_hls_0}
S.~Gupta \emph{et~al.}, ``{SPARK}: a high-level synthesis framework for
  applying parallelizing compiler transformations,'' in \emph{VLSID}, 2003.

\bibitem{spark_optimizing_hls_1}
------, ``Coordinated parallelizing compiler optimizations and high-level
  synthesis,'' \emph{TODAES}, 2004.

\bibitem{merlin_compiler}
J.~Cong \emph{et~al.}, ``Source-to-source optimization for {HLS},'' in
  \emph{{FPGAs} for Software Programmers}, 2016.

\bibitem{polyhedral_hls_1}
L.-N. Pouchet \emph{et~al.}, ``Polyhedral-based data reuse optimization for
  configurable computing,'' in \emph{FPGA}, 2013.

\bibitem{halide_fpga}
J.~Pu \emph{et~al.}, ``Programming heterogeneous systems from an image
  processing {DSL},'' \emph{TACO}, 2017.

\bibitem{hetereohalide}
J.~Li \emph{et~al.}, ``{HeteroHalide}: From image processing {DSL} to efficient
  {FPGA} acceleration,'' in \emph{FPGA}, 2020.

\bibitem{parallel_patterns}
D.~{Koeplinger} \emph{et~al.}, ``Automatic generation of efficient accelerators
  for reconfigurable hardware,'' in \emph{ISCA}, 2016.

\bibitem{frost}
E.~D. Sozzo \emph{et~al.}, ``A common backend for hardware acceleration on
  {FPGA},'' in \emph{ICCD}, 2017.

\bibitem{chisel}
J.~Bachrach \emph{et~al.}, ``{Chisel}: constructing hardware in a scala
  embedded language,'' in \emph{DAC}, 2012.

\bibitem{firrtl}
A.~Izraelevitz \emph{et~al.}, ``Reusability is {FIRRTL} ground: Hardware
  construction languages, compiler frameworks, and transformations,'' in
  \emph{ICCAD}, 2017.

\bibitem{maxcompiler}
{Maxeler Technologies}, ``Programming {MPC} systems (white paper),'' 2013.

\bibitem{intel_opencl_best_practices}
{Intel {FPGA} {SDK} for {OpenCL} {Pro} {Edition} Best Practices Guide,
  UG-OCL003, revision 2020.04.1}. Accessed May 15, 2020.

\end{thebibliography}

\vspace{-4em}
\begin{IEEEbiographynophoto}{Johannes de Fine Licht}
\scriptsize
is a PhD student at ETH Zurich. His research topics revolve around
spatial computing systems in HPC, and include programming models, applications,
libraries, and enhancing programmer productivity.
\end{IEEEbiographynophoto}
\vspace{-4em}
\begin{IEEEbiographynophoto}{Maciej Besta}
\scriptsize
is a PhD student at ETH Zurich. His research focuses on understanding and
accelerating large-scale irregular graph processing in any type of setting
and workload.
\end{IEEEbiographynophoto}
\vspace{-4em}
\begin{IEEEbiographynophoto}{Simon Meierhans}
\scriptsize
is studying for his MSc degree at ETH Zurich. His interests include randomized
and deterministic algorithm and data structure design.
\end{IEEEbiographynophoto}
\vspace{-4em}
\begin{IEEEbiographynophoto}{Torsten Hoefler}
\scriptsize
is a professor at ETH Zurich, where he leads the Scalable Parallel Computing
Lab. His research aims at understanding performance of parallel computing
systems ranging from parallel computer architecture through parallel programming
to parallel algorithms.
\end{IEEEbiographynophoto}

\end{document}